\documentclass[journal]{IEEEtran}
\usepackage{mathrsfs}
\usepackage{bbm}
\usepackage{amssymb}
\usepackage{bbding}
\usepackage{threeparttable}
\usepackage{booktabs}
\usepackage[mathcal]{euscript}

\usepackage{psfrag,calc,url,bm}

\usepackage{cite}

\usepackage{graphicx}
\usepackage{psfrag}

\usepackage{subcaption}
\usepackage{ulem} 
\usepackage{url}
\usepackage{cite}
\usepackage{stfloats}

\usepackage{amsmath}

\usepackage{float}

\usepackage{algorithmic}

\usepackage[ruled,linesnumbered,vlined]{algorithm2e}

\usepackage{color}

\usepackage{boxedminipage}

\usepackage{amsthm}

\usepackage{multirow}

\usepackage{setspace}

\usepackage{soul}
\usepackage{epstopdf}
\usepackage{ulem}
\allowdisplaybreaks[3]

\begin{document}

\title{GNN-Enabled Robust Hybrid Beamforming with Score-Based CSI Generation and Denoising}
\author{Yuhang Li, Yang Lu,~\IEEEmembership{Senior Member,~IEEE}, Bo Ai,~\IEEEmembership{Fellow,~IEEE},  Zhiguo Ding,~\IEEEmembership{Fellow,~IEEE}, \\Arumugam Nallanathan,~\IEEEmembership{Fellow,~IEEE}
\thanks{Yuhang Li and Yang Lu  are with the State Key Laboratory of Advanced Rail Autonomous Operation, and also with the School of Computer Science and Technology, Beijing Jiaotong University, Beijing 100044, China (e-mail: 24110137@bjtu.edu.cn; 
 yanglu@bjtu.edu.cn).}
\thanks{Bo Ai is with the School of Electronics and Information Engineering, Beijing Jiaotong University, Beijing 100044, China (e-mail: boai@bjtu.edu.cn).}
\thanks{Zhiguo Ding is with the School of Electrical and Electronic Engineering (EEE), Nanyang Technological University, Singapore 639798 (Zhiguo.ding@ntu.edu.sg).}
\thanks{Arumugam Nallanathan is with the School of Electronic Engineering and Computer Science, Queen Mary University of London, London and also with the Department of Electronic Engineering, Kyung Hee University, Yongin-si, Gyeonggi-do 17104, South Korea (e-mail: a.nallanathan@qmul.ac.uk).}
}

\maketitle

\begin{abstract}
Accurate Channel State Information (CSI) is critical for Hybrid Beamforming (HBF) tasks. However, obtaining high-resolution CSI remains challenging in practical wireless communication systems. To address this issue, we propose to utilize Graph Neural Networks (GNNs) and score-based generative models to enable robust HBF under imperfect CSI conditions. Firstly, we develop the Hybrid Message Graph Attention Network (HMGAT), which updates both node and edge features through node-level and edge-level message passing. Secondly, we design a Bidirectional Encoder Representations from Transformers (BERT)-based Noise Conditional Score Network (NCSN) to learn the distribution of high-resolution CSI, facilitating CSI generation and data augmentation to further improve HMGAT's performance. Finally, we present a Denoising Score Network (DSN) framework and its instantiation, termed DeBERT, which can denoise imperfect CSI under arbitrary channel error levels, thereby supporting robust HBF. Experiments on DeepMIMO urban datasets demonstrate that the proposed models achieve superior generalization, scalability, and robustness across various HBF tasks for both perfect and imperfect CSI.
\end{abstract}

\begin{IEEEkeywords}
Hybrid Beamforming, Graph Neural Networks, Hybrid Message Graph Attention Network, Bidirectional Encoder Representations from Transformers, Noise Conditional Score Network, Denoising Score Network.
\end{IEEEkeywords}

\section{Introduction}
The rapid evolution of future wireless communication systems requires advanced techniques to address increasingly complex signal processing tasks. Recently, Artificial Intelligence (AI) has emerged as a promising tool to optimize wireless communication systems \cite{2019:KB}. Early attempts to apply {Deep Learning} (DL) in wireless communications have mainly relied on Multilayer Perceptrons (MLPs)\cite{2019:Huang} and Convolutional Neural Networks (CNNs)\cite{2019:Elbir}. While MLPs provide strong nonlinear approximation capabilities  and CNNs are effective in capturing local spatial correlations, both models treat the input data as vectors or regular grids. As a result, they fail to fully exploit the inherent topological structures of wireless communication networks, such as user-Base Station (BS) associations and inter-user interference patterns\cite{2021:yifei}. This limitation becomes particularly critical in large-scale systems, where ignoring the relational structure among entities often leads to suboptimal performance and poor scalability\cite{2023:yifei}.

Among various AI models, Graph Neural Networks (GNNs) have attracted significant attention due to their ability to represent and process irregular graph-structured data\cite{2021:HeShiwen}. By explicitly modeling the interactions between nodes and edges, GNNs naturally capture the underlying topology of communication networks, making them suitable for a wide range of wireless communication tasks such as power allocation\cite{2022:GuoJia}, link scheduling\cite{2022:HeShiwen}, and Hybrid Beamforming (HBF)\cite{2025:yuhang}. Most existing GNN methods for wireless communication systems update only node features via message passing, while neglecting the use of edge features to assist aggregation\cite{11339899}. Although some recent communication-oriented studies incorporate edge information \cite{2022:Tianrui,2023:Shiwen,2021:Mengyuan}, explicit edge feature modeling and updating remain underexplored. Notably, in other GNN-enabled applications\cite{NENN,EHGNN,CensNet}, edge-level message-passing mechanisms have been validated to effectively enhance GNNs. Moreover, in practical wireless communication systems, obtaining perfect Channel State Information (CSI) is challenging, which leads to only imperfect CSI being available at BSs. Existing robust methods typically train models with imperfect CSI to improve resilience\cite{DAQE:2021,DAQE:2024,DQNN,MLP_TL,UMGNN}. 
{However, training directly with imperfect CSI can lead to a statistical characteristic mismatch between the training input CSI and practical CSI, and overlook inherent estimation errors. This may  mislead the model into fitting noisy CSI rather than capturing the practical CSI properties. As a result, the effective expressiveness and generalization performance of the model can be degraded, especially when deployed under high-quality CSI conditions.} This highlights the need for approaches that enhance robustness without sacrificing expressive power.

To tackle the aforementioned challenges, we propose a novel GNN model for sum-rate maximization in HBF tasks, together with two frameworks based on score-based (diffusion) generative models. Specifically, these two frameworks target CSI generation and CSI denoising, respectively. To the best of our knowledge, this is the first work that leverages generative models for both data augmentation and CSI denoising. The main contributions of this paper are summarized as follows:
\begin{itemize}
    \item We propose the Hybrid Message Graph Attention Network (HMGAT), which employs joint node-level and edge-level message passing mechanisms to update node and edge features, respectively. Specifically, for each directed edge, we treat edges sharing the same source node as its neighbors, enabling edge feature aggregation and explicit edge feature updating. This design enhances the representation capacity of GNNs for graph optimization tasks that inherently involve both nodes and edges.
    
    \item We propose a Bidirectional Encoder Representations from Transformers (BERT)-based Noise Conditional Score Network (NCSN) to learn the distribution of high-resolution CSI samples{\footnote{{Since perfect CSI is unachievable, high-resolution CSI samples (obtained by sophisticated channel estimation methods with high pilot overhead) can be used as the ground truth for both CSI generation and denoising.}}} by estimating the corresponding  score function. The BERT-based NCSN enables the generation of high-resolution CSI samples from noisy samples, which can be utilized for data augmentation to boost the generalization performance of HMGAT.
    
    \item We propose a BERT-based Denoising Score Network (DSN), termed DeBERT, for  denoising imperfect CSI under arbitrary channel error levels. The BERT-based DSN can refine imperfect CSI to a high-quality estimation of the perfect CSI, which can be utilized to enhance  the robustness of the HMGAT against CSI errors. 
    
    \item We conduct  experiments on the publicly available DeepMIMO urban datasets, covering three cities and two antenna configurations, resulting in six scenarios in total. For comparison, we  include 14 baselines from existing literature. Numerical results demonstrate that the HMGAT exhibits strong generalization and scalability for HBF design tasks across different scenarios. The proposed BERT-based NCSN effectively improves HMGAT's performance by incorporating the generated CSI samples into the training set. Moreover, DeBERT is capable of denoising imperfect CSI under varying error levels, significantly improving HMGAT's robustness while preserving its expressive power.

\end{itemize}

The remainder of this paper is organized as follows: Section II reviews related works on GNNs, generative models, and robustness-enhancing techniques for wireless communications. Section III introduces the system model, the sum-rate HBF problem as well as its graph optimization formulation, and the imperfect CSI setting. Section IV presents the proposed HMGAT for sum-rate HBF design.  Section V describes the score-based CSI generation  framework and the BERT-based NCSN. Section VI details the DSN and DeBERT for CSI denoising. Section VII provides extensive numerical results and analysis. Finally, Section VIII concludes the paper.

\section{Related Work}
\subsection{GNNs for Beamforming}

GNNs have been increasingly applied to beamforming design in wireless communication systems due to their ability to capture the underlying graph structure of multi-user networks. Early works focused on classical beamforming tasks. For instance, an unsupervised GNN was proposed in \cite{2021:Tianrui} for D2D systems, which learns primal power and dual variables and then converts them to beamforming vectors, reducing the problem dimensionality while achieving strong generalization. In multi-user Multi-Input-Single-Output (MISO) systems, GNNs have been extended to jointly optimize user scheduling and beamforming. The J-USBF algorithm \cite{2023:Shiwen} combined graph-based learning with analytical beamforming solutions, delivering near-optimal performance with high computational efficiency and adaptability to dynamic scenarios. To further improve energy efficiency, Graph Attention Networks (GATs) have been employed to embed node features reflecting inter-link interference \cite{2023:yuhangEE}, while the complex residual GAT \cite{GAT} directly maps CSI to beamforming vectors using attention  and residual connections, enabling fast computation and scalability.

Building on these classical beamforming approaches, recent studies have applied GNNs to HBF in Millimeter-Wave (mmWave) multi-antenna systems. For example, P-PONet \cite{2024:YangJunyi} learned downlink multi-user analog and digital hybrid precoders from uplink sounding signals in an end-to-end manner without explicit channel estimation. Similarly, \cite{2024:WangRuiming} modeled BSs and users as different types of nodes and employed a GNN to map uplink signals to downlink HBF. To further exploit structural properties, \cite{2022:LiuShengjie} proposed a GNN leveraging phase invariance, while \cite{2024:LiuShengjie} introduced a multidimensional GNN that updates hidden representations of hyper-edges to alleviate information loss. Most recently, hybrid analog and digital beamforming has been addressed using both homogeneous and heterogeneous GATs \cite{2025:yuhang}, improving adaptability and representation capability across dynamic network scenarios. 

Despite these advances, most existing GNN-based beamforming methods focus primarily on updating node features, while edge features are either used passively for aggregation or not updated explicitly. As a result, edge features fail to capture the full graph-level information, which limits the expressive power of the models for tasks that inherently involve both nodes and edges. This observation motivates the incorporation of joint node-level and edge-level message passing mechanisms to fully leverage the structural information in GNN-enabled signal processing designs.

\subsection{Node-level and Edge-level Message Passing Mechanisms}

Several works have introduced architectures that incorporate joint node-level and edge-level message passing to capture richer structural information for graphs. NENN \cite{NENN} alternated node-level and edge-level attention layers, allowing node and edge embeddings to mutually reinforce each other. EHGNN \cite{EHGNN} transformed edges into nodes via a dual hypergraph, enabling message passing on edges and producing holistic edge representations. CensNet \cite{CensNet} employed a linear graph transformation to interchange node and edge roles and applied novel graph convolutions to propagate both types of features. 

These studies underscore  the importance of explicitly updating edge features, thereby  providing valuable inspiration for communication-oriented GNN models, since wireless networks are inherently representable as graphs consisting of nodes and edges.

\subsection{Robustness-Enhancing Methods in DL-Based Communication Models}

Existing  robustness-enhancing approaches  mainly rely on training strategies, which can be categorized into two types. The first type, i.e., noise-augmented methods \cite{DAQE:2021,DAQE:2024,DQNN}, generates noisy CSI samples during training and pairs them with the corresponding perfect CSI to enhance model robustness. For instance, {Data Augmentation based Quantile Estimation} (DAQE) \cite{DAQE:2021,DAQE:2024} employed unsupervised or data-augmented DL to handle channel uncertainties, achieving higher data rates and efficiency compared to traditional optimization methods, while a Deep Quantization Neural Network (DQNN) \cite{DQNN} was proposed to tackle imperfect CSI for Reconfigurable Intelligent Surface (RIS)-aided systems. The second type, i.e., Robust Training methods \cite{MLP_TL,UMGNN}, directly injects noise into training CSI to boost  resilience, such as \cite{MLP_TL} for large-scale arrays and \cite{UMGNN} for joint uni-cast and multi-cast beamforming. 

Although the two strategies can enhance robustness, they often compromise the DL model’s inherent expressive power, resulting in a trade-off between accuracy and resilience under imperfect CSI conditions.

\subsection{Generative Models for Wireless Communications}

Generative models have recently attracted attention in wireless communications. One line of research  focused on semantic communications, where diffusion-based models were used for tasks such as denoising, data reconstruction, and content generation \cite{Semantic}. These models, often combined with compression techniques such as variational autoencoders, deliver improved signal fidelity and semantic performance metrics under bandwidth constraints. Another line targeted channel estimation, where score-based generative models learned the gradient of the CSI distribution to iteratively refine imperfect CSI estimations\cite{UNet}. Such models demonstrated robust performance both in- and out-of-distribution, achieving high-fidelity channel estimation even with reduced pilot density. 

Despite these studies, generative models for wireless communications have mostly been confined to limited application scenarios, leaving broader applications to  be explored, such as data augmentation and robust transmission enhancement.

\section{Problem Definition}

In this section, we first present the mmWave communication system model, and formulate the sum-rate maximization problem. Then, we convert the problem into an equivalent graph optimization problem. Finally, we detail the imperfect CSI setting and the associated tasks.

\subsection{System Model and Sum-Rate HBF Design}

Consider a downlink Multi-User MISO (MU-MISO) mmWave system where a BS serves $K$ single-antenna users {\cite{2025:yuhang}}.  The BS is equipped with $N_{\rm T}$ antennas and $N_{\rm F}$ RF chains under the assumption that $N_{\rm T} \ge N_{\rm F} \ge K$. In this case, the fully digital beamforming is infeasible  due to limited number of RF chains. For cost efficiency, the BS adopts the two-stage hybrid digital and analog beamforming architecture. Besides, we assume that each user is served by a single stream, and the BS selects $K$ out of the $N_{\rm F}$ RF chains to serve the $K$ users.

Denote ${s_k}$ as the desired information symbol of the user $k$ with ${\mathbb E}\{|{s_k}|^2\}={\beta_k}$, where ${\beta_k}\ge 0$ denotes the corresponding allocated transmit power. Then, the superposed signal transmitted by the BS is given by
\begin{flalign}
{\bf{x}} = \sum\nolimits_{k = 1}^K {{{\bf P}_{\rm RF}}{{\bf p}_{{\rm BB},k}}{s_k}}, 
\end{flalign}
where ${{\bf P}_{\rm RF}}\in {\mathbb C}^{N_{\rm T} \times K}$ and ${{\bf p}_{{\rm BB},k}}\in {\mathbb C}^{K}$ denote the analog (also refer to radio frequency) and digital (also refer to baseband) precoders, respectively. 

The received signal at the user $k$ is expressed as
\begin{flalign}\label{rec_noise}
{y_k} = {\bf h}_k^H{\bf x} + n_k = &{\bf{h}}_k^H{{\bf{P}}_{{\rm{RF}}}}{{\bf{p}}_{{\rm{BB}},k}} {s_k} + \nonumber\\
&\sum\nolimits_{j = 1,j \ne k}^K {{\bf{h}}_k^H{{\bf{P}}_{{\rm{RF}}}}{{\bf{p}}_{{\rm{BB}},j}} {s_j}}  + {n_k}, 
\end{flalign}
where $n_k\sim {\cal CN}(0,\sigma^2_k)$ denotes the Additive White Gaussian (AWGN) noise. The achievable rate of the user $k$ is given by
\begin{flalign}
R_k&\left(\mathbf{h}_{k}, {{{\bf{P}}_{{\rm{RF}}}},\left\{ {{{\bf{p}}_{{\rm{BB}},i}}},\beta_i \right\}} \right) = \nonumber\\
&{\log _2}\left( {1 + \frac{\beta_i{{{\left| {{\bf{h}}_k^H{{\bf{P}}_{{\rm{RF}}}}{{\bf{p}}_{{\rm{BB}},k}}} \right|}^2}}}{{\sum\nolimits_{j = 1,j \ne k}^K {{{\beta_j\left| {{\bf{h}}_k^H{{\bf{P}}_{{\rm{RF}}}}{{\bf{p}}_{{\rm{BB}},j}}} \right|}^2}}  + {\sigma_k^2}}}} \right).
\end{flalign}

Our goal is to maximize the system sum rate by finding the optimal  ${{\bf{P}}^{\star}_{{\rm{RF}}}}$, $\{ {{{\bf{p}}^{\star}_{{\rm{BB}},i}}} \}$ and $\{\beta^{\star}_i\}$ that solve
\begin{subequations}\label{po1}
\begin{align}
& \left\{{{{\bf{P}}^{\star}_{{\rm{RF}}}},\{ {{{\bf{p}}^{\star}_{{\rm{BB}},i}}},\beta^{\star}_i \}}\right\}  \nonumber= \\
&\arg \max \sum\nolimits_{k = 1}^K {{R_k}\left( {{{\bf{P}}_{{\rm{RF}}}},\left\{ {{{\bf{p}}_{{\rm{BB}},i}}},\beta_i \right\}} \right)} \label{p0:a}\\
{\rm s.t.}~& \left[{{\bf{P}}_{{\rm{RF}}}}\right]_{:,k} \in {\cal F},~\forall k, \label{c_prf}\\
& \sum\nolimits_{k = 1}^K {{\beta _k}}  \le {P_{\max }}, \label{power_con} \\
& \left\|{{\bf{P}}_{{\rm{RF}}}}{{\bf{p}}_{{\rm{BB}},k}}\right\|_2^2 = 1,~\forall k \label{norm_con} , 
\end{align}
\end{subequations}
where $\cal F$ denotes the RF beamforming codebook which satisfies $|[{{\bf P}_{\rm RF}}]_{t,k}|^2=1/N_{\rm T}$, and $P_{\max}$ denotes the power budget of the BS.


\subsection{Graph Representation and Graph Optimization}


In solve Problem \eqref{po1} using GNNs, we need to represent the considered system model as a graph and reformulate  Problem \eqref{po1} into its graph optimization version. 

\begin{figure}[h]
{\centering
{\includegraphics[ width=.45\textwidth]{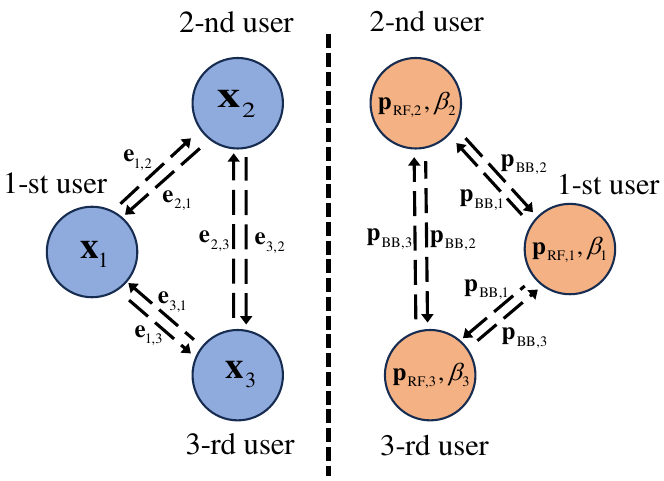}}}
\caption{{Illustration of the graph representation and optimization for a $3$-user case. {Left: CSI-based directed complete graph representation of the MU-MISO system. Right: graph-based HBF optimization with node-level analog/power control and edge-level digital precoding.}}}
\label{fig:graph}
\end{figure}

The MU-MISO mmWave system can be modeled as a homogeneous graph. Specifically, the graph is  directed and complete, denoted as ${\cal G}=({\cal V},{\cal E})$, as shown in the left-hand side of Fig.~\ref{fig:graph}, where ${\cal V}$ represents the set of nodes and ${\cal E}$ represents the set of edges. In this graph, the node $i$ corresponds to the $i$-th user , and its associated node feature is the CSI of the user, defined as  
\begin{flalign}
\mathbf{x}_i \triangleq [\Re({\mathbf{h}_i}^{\top}), \Im({\mathbf{h}_i}^{\top})]^{\top} \in \mathbb{R}^{2N_{\rm T}}.
\end{flalign}
For a directed edge from the node $i$ to the node $j$, the corresponding edge feature is given by  
\begin{flalign}
\mathbf{e}_{i,j} \triangleq \big[\Re(\widetilde{\mathbf{e}}_{i,j}^{\top}), \Im(\widetilde{\mathbf{e}}_{i,j}^{\top})\big]^{\top},
\end{flalign}
where $\widetilde{\mathbf{e}}_{i,j}$ can be constructed from the users' CSI. A typical example is to set $\widetilde{\mathbf{e}}_{i,j}$ by
\begin{flalign}
\widetilde{\mathbf{e}}_{i,j} = \big[{\bf h}_{i}^H {\bf h}_{i}, {\bf h}_{i}^H {\bf h}_{j}, {\bf h}_{j}^H {\bf h}_{j}\big]^{\top}.
\end{flalign}

Based on the defined graph, Problem \eqref{po1} can be formulated as a graph optimization task, as illustrated on the right-hand side of Fig.~\ref{fig:graph}. Specifically, the analog precoder design and power allocation are formulated as node-level optimization tasks, while the digital precoder design is formulated as an edge-level optimization task. {By ensuring that the input and output dimensions of both nodes and edges are independent of the number of users \(K\), GNNs operating on the proposed graph representation can naturally generalize to scenarios with varying user counts.}



\subsection{Imperfect CSI Setting}

Obtaining high-resolution CSI $\mathbf{H} \triangleq [\mathbf{h}_1;\cdots; \mathbf{h}_K] \in \mathbb{C}^{N_{{\rm T}}\times K}$ remains challenging. Although $\mathbf{H}$ can be obtained via some complicated channel estimation techniques, it may consume excessive wireless resources, thereby compromising the efficiency of information transmission. In most cases, only the imperfect\footnote{{Notably, the DL-based robust transmission and convex optimization based robust transmission are fundamentally different. The former aims to improve achievable rates by denoising the imperfect CSI, whereas the latter (via the S-procedure and Bernstein-type inequality) focuses on worst-case or outage-constrained performance. }} CSI matrix $\widetilde{\mathbf{H}}\in \mathbb{C}^{N_{{\rm T}}\times K}$ is available at the BS, which is given by     
\begin{flalign}\label{impcsi}
    \widetilde{\mathbf{H}} = \mathbf{H} + \mathbf{E},
\end{flalign}
where $\mathbf{E} \in \mathbb{C}^{N_{\rm T}\times K}$ represents the channel estimation error\cite{DAQE:2021}. We consider that elements within $\mathbf{E}$ are independent and identically distributed (i.i.d.) and follow a complex Gaussian distribution with $\delta^2_{\rm E}$ being the error variance. {Here, we assume that $\delta^2_{\rm E}$ is available at the BS, which can be roughly estimated through online statistical approaches or historical CSI measurements~\cite{2020:Jiwook}.} 

The main idea to address the graph optimization problem is to first  recover ${\mathbf{H}}$ from its imperfect observation $\widetilde{{\mathbf{H}}}$, and establish a GNN-based mapping from ${\mathbf{H}}$ to feasible $\{{{\bf{P}}_{{\rm{RF}}}},{{{\bf{p}}_{{\rm{BB}},i}}},\beta_i \}$ that maximizes the sum rate. Specifically, there exist three technical tasks:
\begin{itemize}
    \item {Building a powerful GNN model.} A GNN model is required to yield near-optimal solutions with fast inference speed and good scalability to dynamic wireless environments. 
    \item {Augmenting limited high-resolution CSI samples.} The data-driven GNN training  heavily relies on sufficient high-resolution CSI samples, which however, are available only in limited volumes. Therefore, a CSI generation model is required to produce additional high-resolution CSI samples for data augmentation.
    \item {Denoising imperfect CSI observations.} Given an imperfect CSI observation $\widetilde{{\mathbf{H}}}$, a denoising model is required to refine $\widetilde{{\mathbf{H}}}$  into a high-quality input to the GNN model based on the prior knowledge of $\delta^2_{\rm E}$. 
\end{itemize}

\begin{figure}[t]
{\centering
{\includegraphics[ width=.49\textwidth]{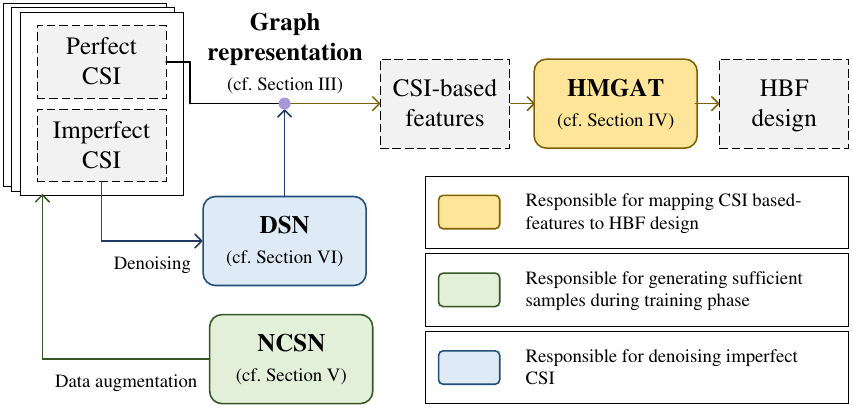}}}
\caption{Illustration of the connections of the three  proposed models.}
\label{fig:illustration}
\end{figure}

In the following  three sections, we respectively elaborate on our proposed GNN model, CSI generation model, and CSI denoising model. For clarity, we use Fig. \ref{fig:illustration} to illustrate the connections of the three  proposed models



\section{Hybrid Beamforming Design via\\Hybrid Message Graph Attention Network}
To simultaneously address the HBF graph optimization tasks on both nodes and edges, we propose the HMGAT. Specifically, HMGAT consists of multiple Hybrid Message Graph Attention Layers (HMGALs), a feature decoding block, and a constrained output layer. These HMGALs serve as the key components for enhancing the HMGAT's expressive power, where node and edge features are updated separately through two rounds of message passing in each layer. Notably, HMGAT takes as input the CSI-based node and edge features, where the CSI underlying these features can be perfect or imperfect. For the case of imperfect CSI, DCN (cf. Section VI) is employed as a front-end module of HMGAT to denoise it.

\subsection{Hybrid Message Graph Attention Layer}
\begin{figure}[h]
{\centering
{\includegraphics[ width=.45\textwidth]{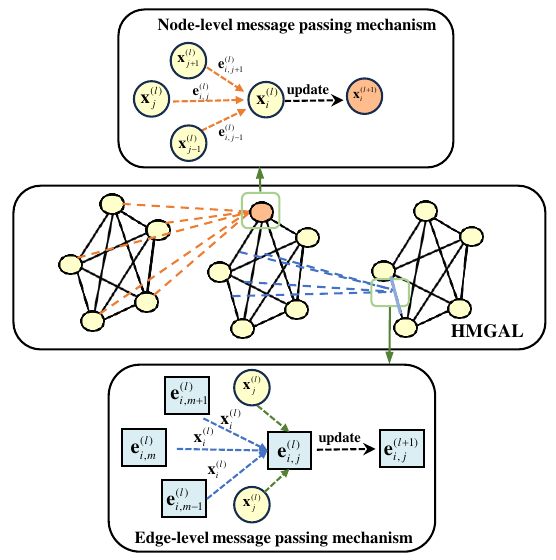}}}
\caption{Illustration of node-level and edge-level message passing mechanisms of HMGAL.} 
\label{fig:hmgal}
\end{figure}

The HMGAL updates node and edge features by performing separate message passing mechanisms at the node and edge levels, ensuring that both node and edge features capture sufficient global graph information for downstream tasks. The node-level and edge-level message passing mechanisms of HMGAL are illustrated in Fig.~\ref{fig:hmgal}.

\subsubsection{Node-Level Message Passing Mechanism}  
For node $i$, its feature is updated by aggregating the features of its neighboring nodes through a multi-head attention mechanism. 
The update process with $M$ attention heads is expressed as
\begin{flalign}
\mathbf{x}^{(l)}_i =
\frac{1}{M} \sum_{m=1}^{M} 
\Bigg( \sum_{j \in \mathcal{N}(i) \cup \{i\}} 
\alpha_{i,j,m} \, \mathbf{\Theta}^{(l)}_m \, \mathbf{x}^{(l-1)}_j \Bigg)
+ \mathbf{\Theta}^{(l)}_{\rm R}\mathbf{x}^{(l-1)}_i,
\end{flalign}
where $\mathbf{x}^{(l)}_i \in \mathbb{R}^{F^{(l)}}$ denotes the feature of node $i$ at the $l$-th layer with ${F^{(l)}}$ being the corresponding dimension, $\mathbf{x}^{(l-1)}_j \in \mathbb{R}^{F^{(l-1)}}$ denotes the feature of node $j$ from the previous layer, $\mathcal{N}(i)$ represents the set of neighbors of node $i$, $\alpha_{i,j,m} \in \mathbb{R}$ is the attention coefficient from the  node $j$ to the node $i$ in the $m$-th head, $\mathbf{\Theta}^{(l)}_m \in \mathbb{R}^{F^{(l)} \times F^{(l-1)}}$ denotes the learnable parameters for the $m$-th attention head at the $l$-th layer, and $\mathbf{\Theta}^{(l)}_{\rm R} \in \mathbb{R}^{F^{(l)} \times F^{(l-1)}}$ denotes the learnable parameters for the residual connection.

The attention coefficients are computed based on both node and edge features, which is given by
\begin{flalign}
\alpha_{i,j,m} = \frac{\exp(e_{i,j,m})}{\sum_{n \in \mathcal{N}(i) \cup \{i\}} \exp(e_{i,n,k})},
\end{flalign}
where $e_{i,j,m}$ represents the unnormalized attention score between the node $i$ and the node $j$, which is computed by
\begin{flalign}
&e_{i,j,m} = \\
&\operatorname{ LeakyReLU}\Big( \mathbf{a}_m^{\top} \big[ \mathbf{\Theta}^{(l)}_m \mathbf{x}^{(l-1)}_i \, \Vert \, \mathbf{\Theta}^{(l)}_m \mathbf{x}^{(l-1)}_j \, \Vert \, \mathbf{\Phi}^{(l)}_m \mathbf{e}^{(l-1)}_{i,j} \big] \Big),\nonumber
\end{flalign}
where $\mathbf{e}^{(l-1)}_{i,j} \in \mathbb{R}^{D^{(l-1)}}$ denotes the edge feature with $D^{(l-1)}$  being the corresponding dimension, $\mathbf{\Phi}^{(l)}_m \in \mathbb{R}^{D^{(l)} \times D^{(l-1)}}$ denotes the learnable parameters for transforming the edge features, $\mathbf{a}_m \in \mathbb{R}^{2F^{(l)} +D^{(l)} }$ denotes the learnable attention parameter, and $\operatorname{ LeakyReLU}(\cdot)$ denotes the LeakyReLU activation function.


\subsubsection{{Edge-Level Message Passing Mechanism}}  
For a directed edge, its feature is updated by aggregating information from other edges that share the identical source. We implement the message passing for edge features through a multi-head attention mechanism that leverages both edge and node features. The edge feature update process with $M$ attention heads is expressed as
\begin{flalign}
 \mathbf{e}^{(l)}_{i,j} =& 
\frac{1}{M} \sum_{m=1}^{M} 
\Bigg( \sum_{n \in \mathcal{N}(i) \cup \{i\}} 
\beta_{(i,j),(i,n),m} \, \mathbf{\widehat{\Phi}}^{(l)}_m \, \mathbf{e}^{(l-1)}_{i,n} \Bigg)+\nonumber
 \\ &  \mathbf{\widehat{\Phi}}^{(l)}_{\rm R} \mathbf{e}^{(l-1)}_{i,j} + \mathbf{\Theta}^{(l)}_{\rm N}\left[ \mathbf{x}^{(l-1)}_i \, \Vert  \,\mathbf{x}^{(l-1)}_j\right],
\end{flalign}
where $\beta_{(i,j),(i,n),m} \in \mathbb{R}$ denotes the attention weight between two edges, $\mathbf{\widehat{\Phi}}^{(l)}_{m} \in \mathbb{R}^{D^{(l)} \times D^{(l-1)}}$ denotes the learnable parameters for edge feature transformation, $\mathbf{\widehat{\Phi}}^{(l)}_{\rm R} \in \mathbb{R}^{D^{(l)} \times D^{(l-1)}}$ denotes the learnable parameters for the residual connection of edge features, and $\mathbf{\Theta}^{(l)}_{\rm N} \in \mathbb{R}^{D^{(l)} \times (2F^{(l-1)})}$ denotes the learnable parameters used to incorporate the concatenated features of the incident nodes.

The normalized attention coefficient is obtained by
\begin{flalign}
\beta_{(i,j),(i,n),m} = 
\frac{\exp\left(z_{(i,j),(i,n),m}\right)}{\sum_{r \in \mathcal{N}(i) \cup \{i\}} \exp\left(z_{(i,j),(i,r),m}\right)},
\end{flalign}
where $z_{(i,j),(i,n),m}$ represents the unnormalized attention score, which is computed by
\begin{flalign}
&z_{(i,j),(i,n),m} \\
&= \operatorname{ LeakyReLU}\left( \mathbf{b}_m^{\top} \left[ 
\mathbf{\widehat{\Phi}}^{(l)}_m \mathbf{e}^{(l-1)}_{i,j} \,\Vert\, 
\mathbf{\widehat{\Phi}}^{(l)}_m \mathbf{e}^{(l-1)}_{i,n} \,\Vert\,
\mathbf{\widehat{\Theta}}^{(l)}_m \mathbf{x}^{(l-1)}_i \right] \right),\nonumber
\end{flalign}
where  $\mathbf{\widehat{\Phi}}^{(l)}_m \in \mathbb{R}^{D^{(l)} \times D^{(l-1)}}$ denotes the learnable parameters for transforming edge features, $\mathbf{\widehat{\Theta}}^{(l)}_m \in \mathbb{R}^{F^{(l)} \times F^{(l-1)}}$ denotes the learnable parameters for transforming the source node feature, and $\mathbf{b}_m \in \mathbb{R}^{2D^{(l)}+F^{(l)}}$ denotes the learnable attention parameters.

\subsection{Feature Decoding Block}

After the processes of $L$ HMGALs, we reach the updated node features $\{\mathbf{x}_i^{(L)}\in\mathbb{R}^{F^{(L)}}\}$ and edge features $\{\mathbf{e}_{i,j}^{(L)}\in\mathbb{R}^{D^{(L)}}\}$. The feature decoding block maps these real-valued features to the complex-valued hybrid-beamforming variables via MLPs followed by a real-to-complex reshaping.

\subsubsection{Node-level decoding (analog precoder \& power allocation)}
Each node feature $\mathbf{x}_i^{(L)}$ is independently decoded by two MLPs:{one for generating the analog precoder block $\mathbf{P}_{{\rm RF}}$ and one for producing the power coefficient $\beta_i$ for each user.} Concretely
\begin{flalign}
\mathbf{P}_{{\rm RF}} = &\operatorname{Reshape}_1\big([\mathrm{MLP}_{\rm RF}(\mathbf{x}_1^{(L)}),\ldots,\mathrm{MLP}_{\rm RF}(\mathbf{x}_K^{(L)})]\big) \nonumber\\  
&\in \mathbb{C}^{N_{\rm T}\times K},\\
\beta_i =& \mathrm{MLP}_{\rm power}(\mathbf{x}_i^{(L)}) \in \mathbb{R},\label{output:beta}
\end{flalign}
where $\mathrm{MLP}_{\rm RF}(\cdot):\mathbb{R}^{F^{(L)}}\to\mathbb{R}^{2N_{\rm T}}$ maps each node feature to a $2N_{\rm T}$-dimensional real vector (representing the real and imaginary parts of one RF precoder column),  $\operatorname{Reshape}_1(\cdot)$ arranges its inputs into an $N_{\rm T}\times K$ complex matrix, and $\mathrm{MLP}_{\rm power}(\cdot):\mathbb{R}^{F^{(L)}}\to\mathbb{R}$ generates the node-specific power coefficient.

\subsubsection{Edge-level decoding (digital precoder)}
{Each edge feature $\mathbf{e}_{i,j}^{(L)}$ is independently decoded by an MLP to obtain the digital precoder vector $\mathbf{p}_{{\rm BB},i}$ associated with user $i$.} Specifically,
\begin{flalign}
&\mathbf{p}_{{\rm BB},i} =\nonumber \\
& \operatorname{Reshape}_2\big([\mathrm{MLP}_{\rm BB}(\mathbf{e}_{i,1}^{(L)}),\ldots,\mathrm{MLP}_{\rm BB}(\mathbf{e}_{i,K}^{(L)})]\big) 
   \in \mathbb{C}^{K},
\end{flalign}
where $\mathrm{MLP}_{\rm BB}(\cdot):\mathbb{R}^{D^{(L)}}\to\mathbb{R}^{2}$ maps each edge feature to a $2$-dimensional real vector (corresponding to the real and imaginary parts of one entry of the digital precoder), and $\operatorname{Reshape}_2(\cdot)$ arranges its inputs into an $K$-dimensional complex vector.


\subsection{Constrained Output Layer}
To ensure that the decoded variables $\{\mathbf{P}_{\rm RF}, \{\mathbf{p}_{{\rm BB},i}\}, \{\beta_i\}\}$ satisfy the constraints of Problem \eqref{po1}, we introduce a constrained output layer that enforces (\ref{c_prf})–(\ref{norm_con}) through carefully designed post-processing operations.

\subsubsection{Power allocation constraint}  
The constraint  \eqref{power_con} requires that the total transmit power does not exceed $P_{\max}$. To enforce this, the raw outputs $\{\beta_i\}$ (cf \eqref{output:beta}) are first passed through a sigmoid activation $\sigma(\cdot)$ and then normalized as
\begin{flalign}
    \beta_i = \frac{\sigma({\beta}_i) P_{\max}}{\max\left(1, \sum_{j=1}^{K} \sigma({\beta}_j)\right)}. \label{PAct}
\end{flalign}

\subsubsection{Analog precoder constraint}  
The constraint \eqref{c_prf} requires each element of the analog precoder to lie in the feasible set $\mathcal{F}$ (e.g., unit-modulus phase shifters). To satisfy this, the raw analog precoder ${\mathbf{P}}_{\rm RF}$ is normalized element-wise as
\begin{flalign}\label{BBAct}
\mathbf{P}_{{\rm RF}_{(i,j)}} =
\frac{{\mathbf{P}}_{{\rm RF}_{(i,j)}}}{|{\mathbf{P}}_{{\rm RF}_{(i,j)}}| \sqrt{N_{\rm T}}}.
\end{flalign}

\subsubsection{Digital precoder constraint}  
To satisfy the per-stream power normalization constraint \eqref{norm_con}, each column of the digital precoder ${\mathbf{P}}_{\rm BB}$ is normalized with respect to the effective precoder $\mathbf{P}_{\rm RF}$:
\begin{flalign}
\mathbf{P}_{{\rm BB}_{(:,i)}} = 
\frac{{\mathbf{P}}_{{\rm BB}_{(:,i)}}}
{\left\|\mathbf{P}_{\rm RF}{\mathbf{P}}_{{\rm BB}_{(:,i)}}\right\|_2}, \label{BBAct:2}
\end{flalign}
such that the combined precoder $\mathbf{P}_{\rm RF}\mathbf{P}_{\rm BB}$ satisfies the total power constraint.


\subsection{Unsupervised Loss Function}
As the constrained output layer guarantees feasible solution, we adopt an unsupervised objective directly based on the achievable sum rate. The unsupervised learning also alleviate the requirement on ground-truth labels
Specifically, the loss function is defined as
\begin{flalign}
{\cal L}_{\rm HMGAT}\left(\bm \theta\right) 
= - \sum_{k=1}^{K} 
R_k\Big( 
\mathbf{P}_{\rm RF},
\{ \mathbf{p}_{{\rm BB},i}, \beta_i \}_{i=1}^K 
\Big)\left|_{\bm \theta}\right.,
\end{flalign}
where $\bm\theta$ denotes all trainable parameters of HMGAT. 

\section{CSI Generation via \\Noise Conditional Score Network}

This section first introduces the core idea of CSI generation framework leveraging score-based generative modeling, and then proposes a BERT-based NCSN. Note that the proposed model can learn the underlying data distribution of the high-resolution CSI samples and generate additional identically distributed CSI samples. {Therefore, NCSN can serve as a data augmentation module to enhance HMGAT by effectively mitigating the issue of insufficient CSI samples during the training phase.}  


\subsection{CSI Generation Framework}

\subsubsection{Score Function (Network)}

Assume that $\bf H$ is sampled from an unknown distribution $\rho_{\rm data}(\mathbf{H})$\cite{2006:Weichselberger}. As there is no explicit expression of $\rho_{\rm data}(\mathbf{H})$, We introduce a parametric score function $\mathbf{S}_{\boldsymbol{\mu}}(\mathbf{H})$, where $\boldsymbol{\mu}$ denotes the learnable parameters, and define it as  
\begin{flalign}\label{s:f}
\mathbf{S}_{\bm \mu}(\mathbf{H})  = \nabla \log \rho_{\rm data}(\mathbf{H}).
\end{flalign}
Here, \eqref{s:f} also indicates that the score function can be represented by a score (neural) network. Notably, $\mathbf{S}_{\bm \mu}(\mathbf{H})$ can be intuitively interpreted as the direction of the steepest increase in the log-probability density. That is, by following the direction guided by $\mathbf{S}_{\bm \mu}(\mathbf{H})$, a Gaussian random sample can converge toward samples of higher likelihood under $\rho_{\rm data}(\mathbf{H})$ {\cite{2019:Song}}.



\subsubsection{Noise Conditional Score Network}

As mentioned, \(\rho_{\rm data}(\mathbf{H})\) is unavailable. Thus, direct score matching is intractable. Instead, we employ the denoising score matching. 

First, we introduce Gaussian perturbation (noise)\footnote{{Notably, the noise in NCSN and the subsequent DSN differs from the AWGN noise in \eqref{rec_noise}, but instead resembles the channel estimation errors in \eqref{impcsi}. Following the definitions of NCSN and DSN, we retain the use of noise.}} ${\bf Z}\in\mathbb{C}^{N_{\rm T}\times K}$ to ${\bf H}$ as
\begin{flalign}
    \overline{\mathbf{H}} = \mathbf{H} + \mathbf{Z},~\mathbf{Z} \sim \mathcal{CN}({\bf 0},\delta^2\mathbf{I}),
\end{flalign}
where $\delta^2$ represents the power of the noise\footnote{For sufficiently small $\delta^2$, the score function of the perturbed sample $\widetilde{\mathbf{H}}$ can be regarded as a reliable approximation of that of the original sample ${\mathbf{H}}$.}. Then, we derive the perturbed score function as
\begin{flalign}
\nabla \log {\rho}_{\rm {data}}\left(\overline{\mathbf{H}} \left| \mathbf{H}\right.\right)\left|_{\delta^2}\right. = -\frac{\overline{\mathbf{H}} - \mathbf{H}}{\delta^2}.
\end{flalign}

Nevertheless, using a single noise in training may bias the score network toward either high-density or low-density regions of the data.  
To address this imbalance, the NCSN~\cite{2019:Song} perturbs  training samples with a sequence of noises with $L$ power levels \(\{\delta_l\}_{l=1}^L\). Thus, we train an NCSN denoted by $\mathbf{S}_{\bm \mu}(\overline{\mathbf{H}}, l)$ to approximate the perturbed score function with given noise level as
\begin{flalign}
\mathbf{S}_{\bm\mu}\left(\overline{\mathbf{H}},l\right) \approx \nabla \log {\rho}_{\rm {data}}\left(\overline{\mathbf{H}} \left| \mathbf{H}\right.\right)\left|_{\delta_l^2}\right.=-\frac{\overline{\mathbf{H}} - \mathbf{H}}{\delta_l^2}. 
\end{flalign}

The supervised training objective aggregates the denoising losses across noise levels as 
\begin{flalign}\label{NCSN:loss}
\mathcal{L}_{\rm NCSN}\left(\bm\mu\right) =\frac{1}{2L}\sum_{l=1}^L  
\Bigg[ \lambda_l 
\Big\| \mathbf{S}_{\bm\mu}\left(\overline{\mathbf{H}}, l\right) + \frac{1}{\delta_l^2}(\overline{\mathbf{H}} - \mathbf{H}) \Big\|_2^2 \Bigg], 
\end{flalign}
where the weight $\delta_l$ can be set by $\lambda_l=\delta_l^2$ to balance the impcat of each noise scale.  

 Notably, the NCSN enables robust estimation of the desired score function, which is crucial for generating CSI samples.


\subsubsection{Annealed Langevin Dynamics}

We generate CSI samples that are identically distributed to real high-resolution CSI from Gaussian noise matrices via Annealed Langevin Dynamics. 

Let $\mathbf{H}_1^{(0)} \in \mathbb{C}^{K\times N_{\rm T}} \sim \mathcal{CN}(\mathbf{0}, \mathbf{I})$ be an initial random Gaussian noise matrix. We iteratively refine $\mathbf{H}_1^{(0)}$ using the well-trained NCSN $\mathbf{S}_{\bm\mu}(\overline{\mathbf{H}}, l)$ to gradually suppress noise components and approximate  the target CSI matrix. 


We assume that $L$ noise power levels follow a descending schedule, i.e., $\delta^2_1 > \delta^2_2 > \cdots > \delta^2_L$.  For each noise level $\sigma^2_l$, we perform $T$ Langevin iterations with the $t$-th iteration being
\begin{equation}
\mathbf{H}_l^{(t+1)} = \mathbf{H}_l^{(t)} + \frac{\nu_l}{2} \mathbf{S}_{\bm\mu}\left(\mathbf{H}_l^{(t)}, l\right) + \sqrt{\nu_{l}} \mathbf{Z}_l^{(t)},~ \mathbf{Z}_l^{(t)} \sim \mathcal{CN}\left(\mathbf{0}, \mathbf{I}\right),
\end{equation}
where $\nu_l= \epsilon \cdot \delta^2_l/\delta_L^2$  denotes the step size for the $l$-th noise power level, with $\epsilon$ serving as the initial step size.

After all iterations, we obtain  $\mathbf{H}_L^{(T)}$ as the  generated high-resolution CSI
sample.

\subsection{{BERT-based Noise Conditional Score Network}}

This subsection proposes a concrete  instantiation of the NCSN $\mathbf{S}_{\bm\mu}(\overline{\mathbf{H}}, l)$ based on BERT. Specifically, this BERT-based instantiation can map a random Gaussian noise matrix to the score function corresponding to a high-resolution CSI matrix at any noise level.

\begin{figure}[h]
{\centering
{\includegraphics[ width=.49\textwidth]{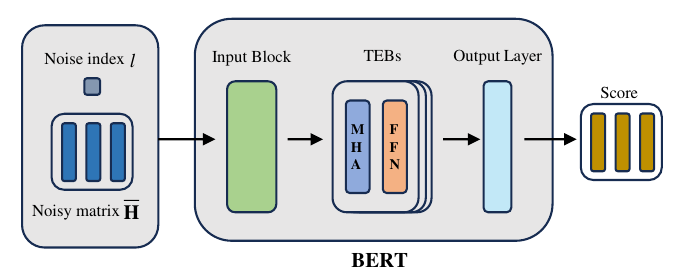}}}
\caption{Illustration of the overall architecture of the BERT-based NCSN.} 
\label{fig:bert_architecture}
\end{figure}

Specifically, the BERT-based NCSN consists of three main modules: an input block, $J$ transformer encoder blocks (TEBs), and an output layer, as illustrated in Fig.~\ref{fig:bert_architecture}.


\subsubsection{Input Block}

The BERT-based NCSN takes as input  the index of noise power level  $l$ and the noisy matrix $\overline{\bf H}$. The input block projects both into a dimension of $D$.

The index  $l$ is encoded into a $D$-dimensional vector via an embedding function $\operatorname{Embed}(\cdot)$. This process is to enable the subsequent TEBs to be conditioned on the noise power level $\delta_l$, which is given by
\begin{equation}
    \mathbf{c}_i = \operatorname{Embed}(i) \in \mathbb{R}^{D}.
\end{equation}

We represent $\overline{\bf H}$ as a real-value matrix $\overline{\bf H}_{\rm R}   \triangleq [\Re({\overline{\bf H}}), \Im(\overline{\bf H})] \in \mathbb{R}^{2N_{\rm T}\times K} $, and project $\overline{\bf H}_{\rm R}$ linearly  to the $D$-dimension space to obtain 
\begin{equation}
    \overline{\mathbf{H}}_{\rm TEB} = \overline{\bf H}_{\rm R}^{\top}\mathbf{W}_{\rm IB} \in \mathbb{R}^{K \times D},
\end{equation}
where $\mathbf{W}_{\rm IB} \in \mathbb{R}^{2N_{\rm T} \times D}$ denotes learnable parameters.

\subsubsection{Transformer Encoder Block}

Each TEB comprises  two sub-blocks, i.e., a Multi-Head self-Attention (MHA) sub-block and a position-wise Feedforward Network (FFN) sub-block. Both sub-blocks are conditioned on $\mathbf{c}_i$ via adaptive modulation and residual gating. The input and output of each TEB reside in  $\mathbb{R}^{K \times D}$.  

The MHA sub-block first applies layer normalization $\text{LN}(\cdot)$ to $\overline{\mathbf{H}}_{\rm TEB}$, and performs the adaptive modulation operation as
\begin{equation}
    \overline{\mathbf{H}}_{\rm AM} =\text{LN}\left(\overline{\mathbf{H}}_{\rm TEB}\right) \odot \left(\mathbf{V}_{\rm scale}\mathbf{c}_i\right) \oplus  \left(\mathbf{V}_{\rm shift}\mathbf{c}_i\right),
\end{equation}
where $\mathbf{V}_{\rm scale}, \mathbf{V}_{\rm shift} \in \mathbb{R}^{D \times D}$ are learnable parameters, {and both the element-wise multiplication $\odot$ and addition $\oplus$ are performed with broadcasting over the feature dimensions.}
 Then,   $\overline{\mathbf{H}}_{\rm AM}$ is projected into queries, keys, and values via $C$ attention heads. Specifically, the $c$-th head gives rise to
\begin{equation}
\mathbf{Q}^{(c)} = \overline{\mathbf{H}}_{\rm AM} \mathbf{W}_{\rm Q}^{(c)},~
\mathbf{K}^{(c)} = \overline{\mathbf{H}}_{\rm AM} \mathbf{W}_{\rm K}^{(c)},~
\mathbf{V}^{(c)} = \overline{\mathbf{H}}_{\rm AM} \mathbf{W}_{\rm V}^{(c)},
\end{equation}
where $\mathbf{W}_{\rm Q}^{(c)}, \mathbf{W}_{\rm K}^{(c)}, \mathbf{W}_{\rm V}^{(c)} \in \mathbb{R}^{D \times D_{\rm h}}$ are learnable projection matrices, with $D_{\rm h} = D/C$ representing the dimension of each attention head. The  output of the $c$-th attention head is
\begin{flalign}
    \mathbf{O}^{(c)} = \text{Softmax}\left(\frac{\mathbf{Q}^{(c)} {\mathbf{K}^{(c)}}^{\top}}{\sqrt{D_{\rm h}}}\right)\mathbf{V}^{(c)} \in \mathbb{R}^{K \times D_{\rm h}},
\end{flalign}
where $\text{Softmax}(\cdot)$ represents the softmax function. By concatenating the outputs of $C$ attention heads via the concatenation operation $\text{Concat}(\cdot)$, followed by a linear projection with a learnable weight matrix of $\mathbf{W}_{\rm O}\in \mathbb{R}^{D \times D}$, we obtain
\begin{flalign}
\overline{\bf H}_{\rm HMA} = \text{Concat}\left(\mathbf{O}^{(1)}, \ldots, \mathbf{O}^{(C)}\right) \mathbf{W}_{\rm O}\in \mathbb{R}^{K \times D}.
\end{flalign}
Finally, a residual gating mechanism governed by $\mathbf{c}_i$ is employed to obtain the output of the MHA sub-block as 
\begin{flalign}\label{out_sub1}
   \overline{\mathbf{H}}_{\rm TEB} \leftarrow \overline{\mathbf{H}}_{\rm TEB} + \big(\overline{\bf H}_{\rm HMA} \odot (\mathbf{V}_{\rm gate}\mathbf{c}_i)\big),
\end{flalign}
where $\mathbf{V}_{\rm gate} \in \mathbb{R}^{D\times D}$ denotes learnable parameters.  

The FFN sub-block first processes  the output of its upstream MHA  sub-block, i.e., $\widetilde{\mathbf{H}}_{\rm TEB}$ in \eqref{out_sub1}, as
\begin{equation}
    \overline{\mathbf{H}}_{\rm AM}' = \text{LN}\left(\overline{\mathbf{H}}_{\rm TEB}\right) \odot (\mathbf{{V}}'_{\rm scale}\mathbf{c}_i) \oplus (\mathbf{{V}}'_{\rm shift}\mathbf{c}_i),
\end{equation}
where $\mathbf{{V}}'_{\rm scale}, \mathbf{{V}}'_{\rm shift} \in \mathbb{R}^{D \times D}$ are learnable parameters. Then, $\widetilde{\mathbf{H}}_{\rm AM}'$ is passed through a two-layer FFN with GELU activation function $\text{GELU}(\cdot)$ as
\begin{equation}
    \overline{\mathbf{H}}_{\rm FFN}' = \text{GELU}\left(\overline{\mathbf{H}}_{\rm AM}' \mathbf{W}_1\right)\mathbf{W}_2,
\end{equation}
where $\mathbf{W}_1 \in \mathbb{R}^{D \times D'}, \mathbf{W}_2 \in \mathbb{R}^{D' \times D}$ are learnable parameters, and $D'$ is the intermediate hidden size. The final residual update is given by
\begin{equation}
    \overline{\mathbf{H}}_{\rm TEB} \leftarrow \overline{\mathbf{H}}_{\rm TEB} + \big(\overline{\mathbf{H}}_{\rm FFN}' \odot (\mathbf{{V}}'_{\rm gate}\mathbf{c}_i)\big)\in\mathbb{R}^{K\times D},
\end{equation}
where ${\bf{V}}'_{\rm gate} \in \mathbb{R}^{D\times D}$ denotes learnable parameters.  


\subsubsection{Output Layer}

The output layer projects $\overline{\mathbf{H}}_{\rm TEB}$ updated by $J$ TEBs back to the CSI dimension:
\begin{equation}
    \mathbf{S}_{\bm\mu}(\overline{\mathbf{H}}, l) \leftarrow \operatorname{Reshape}_1 \big( \overline{\mathbf{H}}_{\rm TEB}^{\top}\mathbf{W}_{\rm OB}\big) \in \mathbb{C}^{ N_{\rm T}\times K},
\end{equation}
where $\mathbf{W}_{\rm OB} \in \mathbb{R}^{D \times 2N_{\rm T}}$ is learnable parameters .

\section{CSI Denoising via \\ Denoising Score Network}
This section first introduces the core idea of the CSI denoising framework including the DSN and its multi-task loss function. Then, the DeBERT is proposed to instantiate the DSN. Note that the well-trained DSN can be utilized to obtain denoised CSI, which is further input into the HMGAT to realize the robust HBF design.



\subsection{CSI Denoising Framework}

The DSN takes as input the imperfect CSI matrix $\widetilde{\bf H}$ (cf. \eqref{impcsi}) and the error variance $\delta_{\rm E}^2$, and enables the denoising of $\widetilde{\bf H}$ to obtain a refined estimation for ${\bf H}$ denoted by $\widehat{\bf H}$.

\begin{flalign}\label{dsn}
\left\{\mathbf{S}_{\rm score}, \bm\Delta, {\bm\eta}\right\} &= {\rm DSN}_{\bm\tau}\left(\widetilde{\mathbf{H}}, \delta_{\rm E}\right),\\
\widehat{\mathbf{H}} &= \widetilde{\mathbf{H}} + {\bm\eta} \odot \bm\Delta\in\mathbb{C}^{N_{\rm T}\times K},\label{dn}
\end{flalign}
where the output of DSN comprises the noise-conditional score $\mathbf{S}_{\rm score}\in\mathbb{C}^{N_{\rm T}\times K}$, the denoising direction $\bm\Delta\in\mathbb{C}^{N_{\rm T}\times K}$, and the step size ${\bm\eta}\in{\mathbb C}^{K}$, and ${\bm \tau}$ represents the learnable parameters of the DSN.

We employ a multi-task training strategy to train the DSN. Specifically, the supervised loss function includes two terms, i.e., score matching error $\mathcal{L}_{\text{score}}({\bm\tau})$ and reconstruction error $\mathcal{L}_{\text{denoise}}({\bm\tau})$, which is given by
\begin{equation}
\mathcal{L}_{\rm DSN}\left({\bm\tau}\right) = \mathcal{L}_{\text{score}} \left({\bm\tau}\right) + \lambda \mathcal{L}_{\text{denoise}} \left({\bm\tau}\right),
\end{equation}
where $\lambda$ is a hyperparameter. The score matching error encourages the DSN to learn the noise-conditional score function of  high-resolution CSI as 
\begin{flalign}\label{l_score}
\mathcal{L}_{\text{score}} \left({\bm\tau}\right)= \delta_{\rm E}^ 2\left\|\mathbf{S}_{\rm score} + \frac{\widetilde{\mathbf{H}} - \mathbf{H}}{\delta_{\rm E}^2}\right\|_2^2.
\end{flalign}
The reconstruction error evaluates the denoising performance for \eqref{dn} as
\begin{equation}\label{l_de}
\mathcal{L}_{\text{denoise}}\left({\bm\tau}\right) = 
\frac{\| \mathbf{H} - \widehat{\mathbf{H}} \|_F}{\| \mathbf{H} \|_F}.
\end{equation}


\subsection{BERT-based Denoising Score Network}

\begin{figure}[h]
{\centering
{\includegraphics[ width=.49\textwidth]{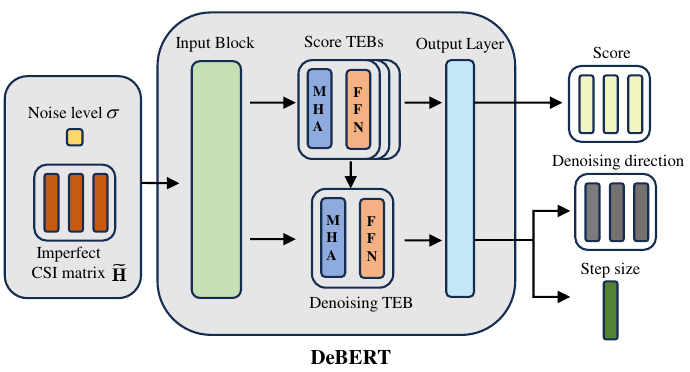}}}
\caption{Illustration of the overall architecture of the DeBERT.} 
\label{fig:debert_architecture}
\end{figure}

This subsection proposes a concrete instantiation of the DNS built upon the BERT architecture, i.e., DeBERT. Specifically, the  DeBERT comprises four core  modules: an input block, $\ddot J$ score TEBs, a denoising TEB, and an output layer, as illustrated in Fig.~\ref{fig:debert_architecture}.

\subsubsection{Input Block} 

The DeBERT takes as input the imperfect CSI matrix   $\widetilde{\bf H}$ (cf. \eqref{impcsi}) and error variance $\delta_{\rm E}^2$ (cf. \eqref{dsn}). The input block projects both into a dimension of $\ddot D$.

The error variance is embedded via a linear transformation, which is given by
\begin{flalign}
    \mathbf{c}_{\delta_{\rm E}} = {\ddot{\bf{w}}}_{\rm IB} \delta_{\rm E} + \ddot{{\bf b}}_{\rm IB}\in \mathbb{R}^{\ddot{D}},    
\end{flalign}
where $\ddot{\mathbf{w}}_{\rm IB} ,\ddot{\mathbf{b}}_{\rm IB} \in \mathbb{R}^{\ddot{D}}$ are learnable parameters.

We represent $\widetilde{\bf H}$ as a real-value matrix $\widetilde{\bf H}_{\rm R}   \triangleq [\Re({\widetilde{\bf H}}), \Im(\widetilde{\bf H})] \in \mathbb{R}^{2N_{\rm T}\times K} $, and project $\widetilde{\bf H}_{\rm R}$ linearly  to the $\ddot D$-dimension space to obtain 
\begin{equation}
    \widetilde{\mathbf{H}}_{\rm TEB} = \widetilde{\bf H}_{\rm R}^{\top}\ddot{\mathbf{W}}_{\rm IB} \in \mathbb{R}^{K \times \ddot{D}},
\end{equation}
where $\ddot{\mathbf{W}}_{\rm IB} \in \mathbb{R}^{2N_{\rm T} \times \ddot{D}}$ denotes learnable parameters. 



\subsubsection{Score TEB and Denoising TEB}

The ${\ddot J}$ score TEBs and the denoising TEB in the DeBERT are implemented analogously  to the standard BERT TEBs (cf. Section V-B), but their functional roles differ. Specifically, the score TEBs enable to yield the noise-conditional score (which is similar to TEBs in the BERT-based NCSN), while the denoising TEB enables to output the denoising direction and the step size. The input and output of each TEB in the DeBERT reside in  $\mathbb{R}^{K \times \ddot{D}}$. 

Each score TEB is defined by
\begin{equation}
\widetilde{\mathbf{H}}_{\rm score}\leftarrow\text{TEB}_{\rm score}\left(\widetilde{\mathbf{H}}_{\rm TEB}, \mathbf{c}_{\delta_{\rm E}}\right) \in \mathbb{R}^{K \times \ddot{D}}.
\end{equation}

The denoising TEB takes as input the concatenation of the original input $\widetilde{\mathbf{H}}_{\rm TEB}$ and the output of the $\ddot J$-th score TEB   $\widetilde{\mathbf{H}}_{\rm Score}$, together with $\mathbf{c}_{\delta_{\rm E}}$, which is given by
\begin{equation}
\widetilde{\mathbf{H}}_{\rm TEB}\leftarrow\text{TEB}_{\rm De }\left(\widetilde{\mathbf{H}}_{\rm TEB} \oplus \widetilde{\mathbf{H}}_{\rm score}, \mathbf{c}_{\delta_{\rm E}}\right)  \in \mathbb{R}^{K \times \ddot{D}}.
\end{equation}


\subsubsection{Output Layer}

The output layer projects the obtained $\widetilde{\mathbf{H}}_{\rm Score}$ to ${\bf S}_{\rm score}$ as
\begin{flalign}
    \mathbf{S}_{\rm score} &= \operatorname{Reshape}_1\left(\widetilde{\mathbf{H}}_{\rm score}\ddot{\mathbf{W}}_{\rm score}\right) , 
\end{flalign}
while projecting the obtained  $\widetilde{\mathbf{H}}_{\rm TEB}$ to $\bm\Delta$ and $\eta$, respectively, as
\begin{flalign}
\bm\Delta &= \operatorname{Reshape}_1\left(\widetilde{\mathbf{H}}_{\rm TEB} \ddot{\mathbf{W}}_{\rm TEB}\right) , \\
{\bm\eta} &= \text{MLP}_{\rm SS}\left(\widetilde{\mathbf{H}}_{\rm TEB}\right),
\end{flalign}
where \(\ddot{\mathbf{W}}_{\rm score}, \ddot{\mathbf{W}}_{\rm TEB} \in \mathbb{R}^{{\ddot D} \times 2N_{\rm T}}\) are learnable parameters, and  $\mathrm{MLP}_{\rm SS}(\cdot):\mathbb{C}^{{N_{\rm T}\times K}}\to\mathbb{C}^{K}$ denotes a  learnable network.


\section{Numerical Results}
This section presents extensive numerical results to validate the  effectiveness and superiority of the proposed methods (i.e., HMGAT, BERT-based NCSN, and DeBERT). First, we evaluate the expressive power and scalability of the proposed HMGAT  under perfect CSI conditions. Then, for the high-resolution CSI generation task, we demonstrate that the CSI samples generated by the BERT-based NCSN can be effectively utilized for data augmentation, which significantly enhances the performance of HMGAT. Moreover, under imperfect CSI conditions, we verify the effectiveness of DeBERT in denoising imperfect CSI, showing its ability to improve the robustness of HMGAT against CSI errors. Finally, we compare and analyze the inference time of the methods under test. 
\subsection{Simulation Setup}
\subsubsection{Simulation Scenario}

The simulation scenarios are derived from the publicly available DeepMIMO ray-tracing dataset. To capture diverse propagation environments, we select BSs from three representative cities, namely New York (NY), Los Angeles (LA), and Chicago (CHI), for the experiments.

Unless otherwise specified, the default system parameters are given as follows: bandwidth $0.5$ GHz, carrier frequency $28$ GHz, and $10$ propagation paths. The corresponding wavelength is approximately $10.7$ mm, and the antenna spacing is set to half-wavelength. The noise power spectral density is assumed to be $-174$ dBm/Hz.



\subsubsection{Dataset}
All DL methods are trained on two datasets with different scales: $(K, N_{\rm T})=(8,16)$ and $(K, N_{\rm T})=(16, 32)$. Each dataset consists of 10,000 samples, which are divided into training, validation, and test sets with a ratio of 8:1:1. During training, the maximum transmit power is set to $P_{\rm{max}} = 1$ W.


\subsubsection{Implementation Detail}
The DL models are trained using the AdamW optimizer, with a learning rate of $5 \times 10^{-4}$, a batch size of 32, and a weight decay of $1 \times 10^{-4}$. The training process runs for 100 epochs with a dropout rate of $10\%$. 


\subsubsection{Baseline}

The baselines used to evaluate the effectiveness of HMGAT are listed as follows:
\begin{itemize}
    \item \textbf{Traditional approaches}:
    \begin{itemize}
        \item [*] \textbf{Phase Zero Forcing (PZF)}~\cite{PZF}: Employs phase-only RF control with baseband ZF precoding.  
        \item [*] \textbf{Alternating Optimizer (AO)}~\cite{yuwei}: Iterative convex-relaxation method implemented with CVX solver.
        \item [*] \textbf{Manifold}~\cite{manifold} : Treats hybrid precoder design as matrix factorization, solved via alternating minimization.
    \end{itemize}
    
    \item \textbf{Conventional DL models}:
    \begin{itemize}
        \item [*] \textbf{CNN}~\cite{CNN}: Applies convolutional layers to extract channel features for HBF design.
        \item [*] \textbf{MLP}~\cite{MLP_TL}:  Uses a fully connected deep neural network with concatenated channel inputs.
    \end{itemize}
    \item \textbf{GNN models}:
    \begin{itemize}
        \item [*] \textbf{GCN}~\cite{GCN}: A basic GNN applying message passing via graph convolution. 
        \item [*] \textbf{GAT}~\cite{GAT}: A GNN with attention-based message passing.
    \end{itemize}
    
    \item \textbf{Node- and edge-GNN models}:
    \begin{itemize}
        \item [*] \textbf{NENN}~\cite{NENN}: GNN jointly modeling nodes and edges via dual-level attention. 
        \item [*] \textbf{EHGNN}~\cite{EHGNN}: GNN using Dual Hypergraph Transformation for richer edge embeddings. 
        \item [*] \textbf{CensNet}~\cite{CensNet}: GNN embedding nodes and edges with edge-node switching and convolution.
    \end{itemize}
\end{itemize}

To demonstrate the effectiveness of the proposed BERT-based NCSN, the following two models are used as baselines, with identical hyperparameter settings applied during both training and sampling.
\begin{itemize}
    \item \textbf{UNet}\cite{UNet}: A CNN-based generative model that utilizes an encoder-decoder architecture with skip connections to capture both local and global structures.
    \item \textbf{DiT}\cite{DiT}: A transformer-based generative model using self-attention to capture long-range dependencies for high-quality generation and denoising.
\end{itemize}

To demonstrate the effectiveness of the proposed DeBERT, we compare it against the following baselines.
\begin{itemize}
    \item \textbf{Noise-Augmented}\cite{DAQE:2021,DAQE:2024,DQNN}: Generates noise samples during training and leverages the corresponding perfect CSI to enhance the model’s robustness.
    \item \textbf{Robust Training}\cite{MLP_TL,UMGNN}: Noise is added to the perfect CSI and directly used for model training, enhancing the model’s robustness.
\end{itemize}


\begin{table*}[htbp]
\centering
\caption{Achievable sum rates of different scenarios.}
\setlength{\tabcolsep}{3pt} 
\begin{tabular}{c|c||c|c|c|c|c|c|c|c|c|c|c}
\hline
{$(K, N_{\rm T})$} & City  & PZF & AO& Manifold & CNN& MLP& GCN& GAT&NENN&CensNet&EHGNN& HMGAT \\
\cline{3-10}
\hline
\hline
\multirow{3}{*}{(8, 16)} & NY&7.09&7.16&3.93&9.93&8.88&9.06&10.70&10.10&11.28&10.12&12.98 \\
& LA &7.16&6.18&4.79&12.83&9.72&9.50&16.32&9.58&17.90&11.96&20.48  \\
& CHI&14.66&14.75&5.42&13.99&13.41&13.02&14.78&14.80&14.59&14.87&19.51  \\
\hline
\multirow{3}{*}{(16, 32)} & NY&8.33&11.94&5.29&11.28&10.03&10.54&11.75&11.90&16.91&15.65&18.78 \\
& LA &10.46&13.50&8.09&15.37&9.21&9.43&24.07&23.33&29.03&13.00&37.61  \\
& CHI&29.64&31.39&10.61&23.34&14.95&14.30&17.05&15.94&17.13&25.37&32.86 \\
\hline
\end{tabular}
\label{table:HMGAT:sumrate}
\end{table*}

\subsection{Effectiveness of HMGAT}
To comprehensively evaluate HMGAT, we assess its performance across different scenarios under perfect CSI, and further examine its scalability under varying transmit power levels and user numbers.
\subsubsection{Performance Comparison} The results in Table~\ref{table:HMGAT:sumrate} show clear differences across method categories. Traditional approaches (PZF, AO and Manifold) achieve the lowest sum rates, e.g., in LA with $(16,32)$, PZF, AO, and Manifold reach only 10.46, 13.50, and 8.09, respectively. Conventional DL models (CNN and MLP) improve performance by exploiting channel features, yet still fluctuate in large-scale settings; for instance, in $(16,32)$ CHI, CNN and MLP achieve sum rates of 23.34 and 14.95, respectively.

GNNs (GCN and GAT) further boost sum rate by leveraging message passing, with GAT performing better in dense scenarios.  Node- and edge-GNNs (NENN, CensNet and EHGNN) gain additional improvements by incorporating edge information. For example, CensNet and EHGNN reach 16.91 and 15.65 in NY with $(16,32)$, and 29.03 and 13.00 in LA with $(16,32)$, demonstrating the benefit of modeling both nodes and edges in large antenna-user configurations.

Most notably, the proposed HMGAT achieves the best performance across all environments and system scales. For example, under the $(8,16)$ setting, the HMGAT improves the sum rate in LA to 20.48, exceeding CensNet by 2.58. Under the $(16,32)$ setting, the HMGAT demonstrates even stronger scalability, achieving 37.61 in LA, which is significantly higher than the best competitor. Similarly, in CHI, the HMGAT reaches 32.86, surpassing AO (31.39) and EHGNN (25.37). These consistent improvements across diverse environments confirm that the HMGAT effectively captures both node- and edge-level dependencies while maintaining scalability under higher user densities and antenna configurations, thereby establishing a new state-of-the-art for the HBF design.

\begin{figure*}[htbp]
	\centering
	\begin{subfigure}{0.32\linewidth}
		\centering
    \includegraphics[width=1\linewidth]{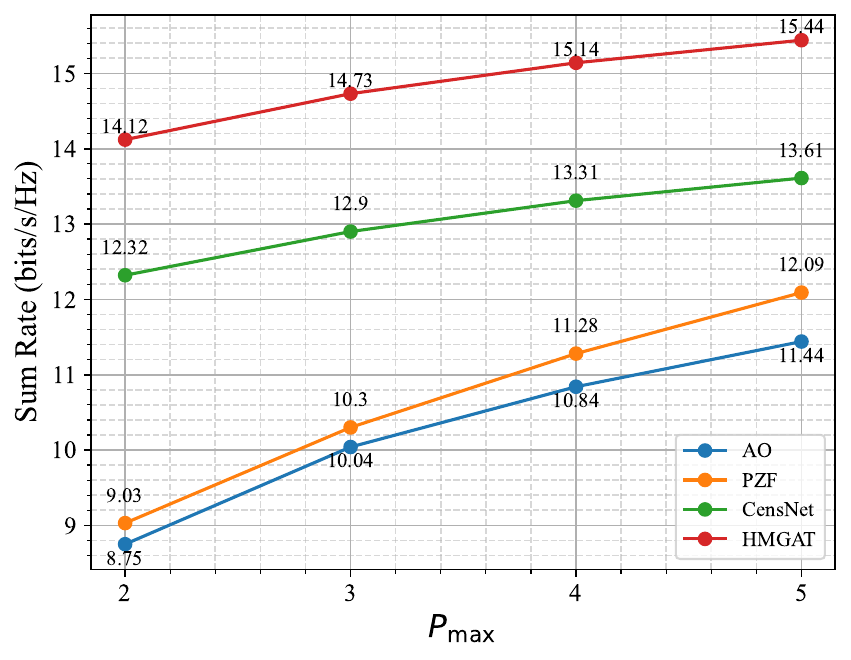}
		\caption{{NY, $(K = 8,N_{\rm T} = 16)$}}
		\label{bs1}
	\end{subfigure}
	\centering
	\begin{subfigure}{0.32\linewidth}
		\centering
		\includegraphics[width=1\linewidth]{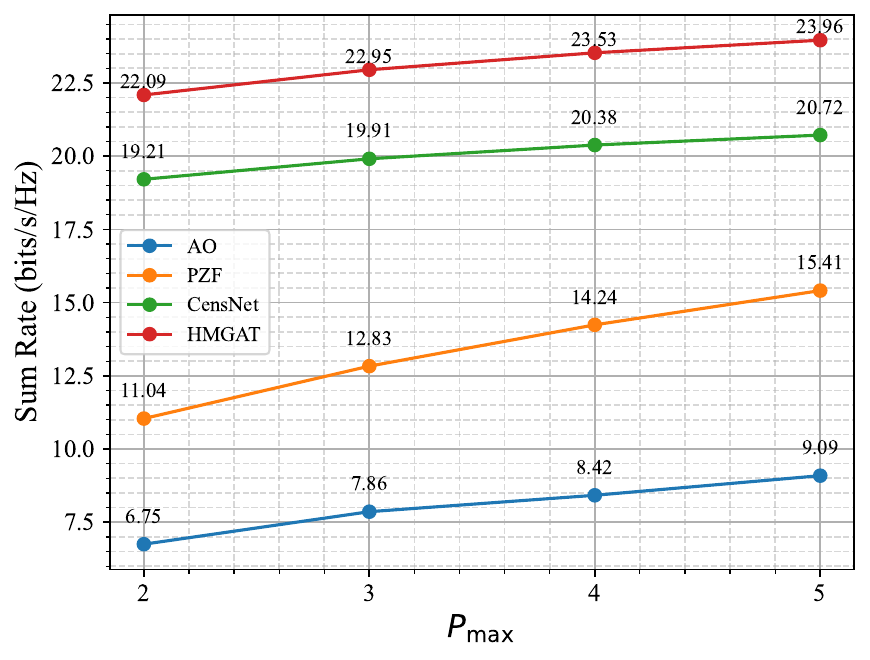}
		\caption{{LA, $(K = 8,N_{\rm T} = 16)$}}
		\label{bs2}
	\end{subfigure}
	\centering
	\begin{subfigure}{0.32\linewidth}
		\centering
		\includegraphics[width=1\linewidth]{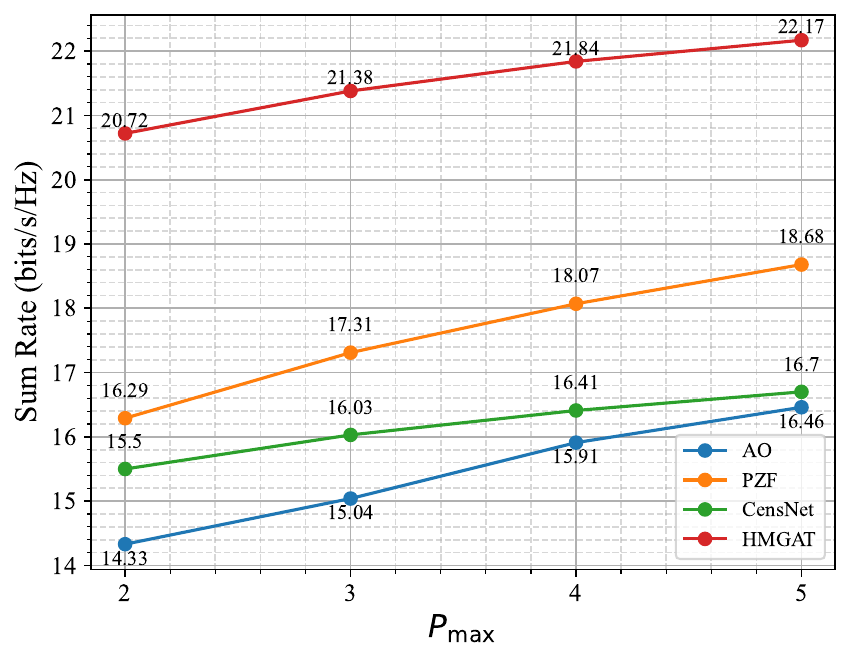}
		\caption{{CHI, $(K = 8,N_{\rm T} = 16)$}}
		\label{bs3}
	\end{subfigure}

	\begin{subfigure}{0.32\linewidth}
		\centering
    	\includegraphics[width=\linewidth]{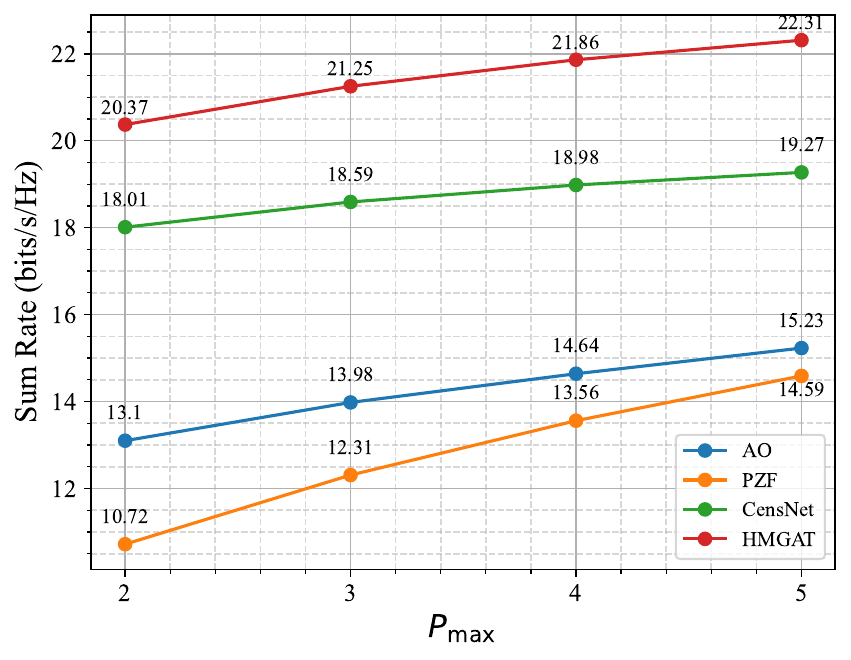}
		\caption{{NY, $(K = 16,N_{\rm T} = 32)$}}
		\label{bs4}
	\end{subfigure}
	\begin{subfigure}{0.32\linewidth}
		\centering
		\includegraphics[width=\linewidth]{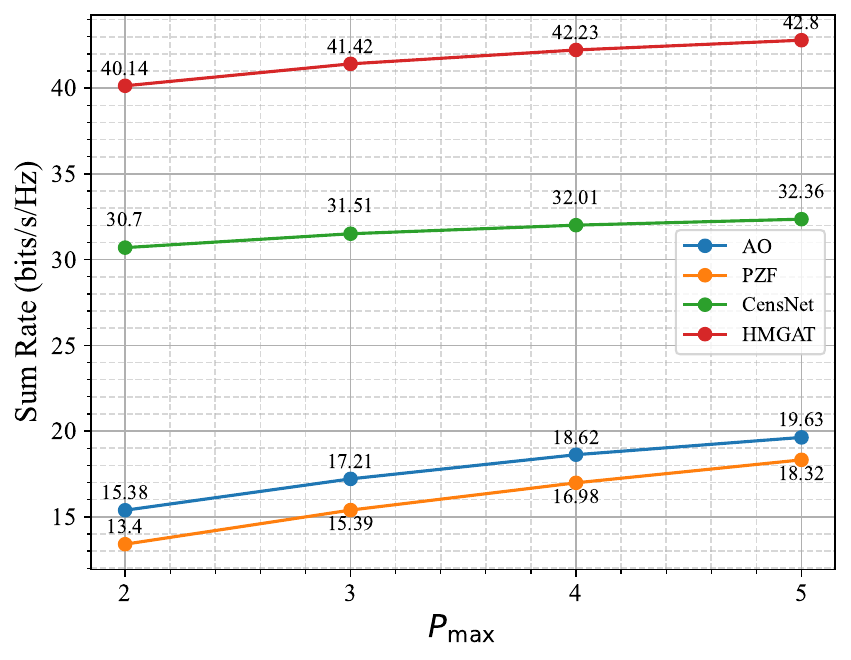}
		\caption{{LA, $(K = 16,N_{\rm T} = 32)$}}
		\label{bs5}
	\end{subfigure}
	\begin{subfigure}{0.32\linewidth}
		\centering
		\includegraphics[width=\linewidth]{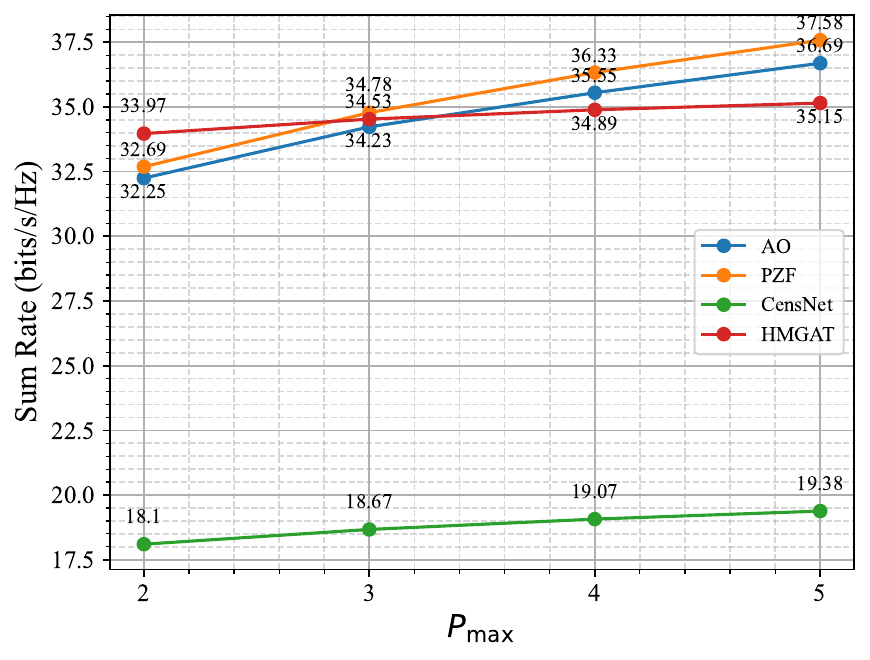}
		\caption{{CHI, $(K = 16,N_{\rm T} = 32)$}}
		\label{bs6}
	\end{subfigure}
	\caption{Sum rates versus power budgets.}
	\label{fig:HMGAT:power}
\end{figure*}

\subsubsection{Scalability Across Different Power Budgets}
Fig.~\ref{fig:HMGAT:power} evaluates the scalability of the top-performing DL models (i.e., HMGAT and CensNet) under different power budgets. We compare the models trained only at $P_{\rm max} = 1$ with  AO and PZF under test scenarios of $P_{\rm max} \in \{2,3,4,5\}$. As observed, the HMGAT demonstrates the best scalability in 5 out of the 6 test scenarios. In CHI with $(16,32)$, HMGAT’s performance at $P_{\rm max} \in \{3,4,5\}$ is slightly lower than AO and PZF, with the maximum gap not exceeding 3. {The reason may be the specific channel conditions and randomly generated user layouts in CHI with $(16,32)$, which can temporarily diminish HMGAT’s advantage.}

\begin{table}[t]
\centering
\small 
\caption{Scalability to different numbers of users.}
\label{tab:sumrate_combined}
\renewcommand{\arraystretch}{1.2}
\setlength{\tabcolsep}{1.5pt} 
\begin{tabular}{c|c||c|c|c|c|c|c|c|c}
\hline
\multirow{2}{*}{City}& \multirow{2}{*}{Model} & \multicolumn{4}{c|}{$(K=8,N_{\rm T}=16)$} & \multicolumn{4}{c}{$(K=16,N_{\rm T}=32)$} \\
\cline{3-10}
 & & 6 & 7 & 9 & 10 & 14 & 15 & 17 & 18 \\
\hline
\hline
\multirow{4}{*}{NY} & AO      & 8.09 & 7.42 & 5.03 & 5.43 & 10.08 & 11.29 & 12.95 & 12.82 \\
 & PZF     & 8.32 & 7.82 & 6.45 & 5.59 & 9.81  & 9.56  & 7.64  & 6.84 \\
 & CensNet & 9.70 & 10.83 & 11.91 & 12.04 & 15.94 & 16.50 & 17.05 & 17.56 \\
 & {\bf HMGAT} & 11.46 & 12.40 & 13.75 & 14.19 & 17.85 & 18.39 & 19.08 & 19.12 \\
\hline
\multirow{4}{*}{LA} & AO      & 8.88 & 7.08 & 5.65 & 5.41 & 12.16 & 12.78 & 15.55 & 11.17 \\
 & PZF     & 11.48 & 9.98 & 7.17 & 6.05 & 12.39 & 11.87 & 9.59  & 9.14 \\
 & CensNet & 16.60 & 17.80 & 18.05 & 17.31 & 27.88 & 28.60 & 29.63 & 29.04 \\
 & {\bf HMGAT} & 18.33 & 19.91 & 21.21 & 21.30 & 35.71 & 37.03 & 38.35 & 39.03 \\
\hline
\multirow{4}{*}{CHI} & AO      & 10.69 & 14.42 & 17.22 & 18.19 & 27.79 & 30.85 & 32.24 & 33.62 \\
 & PZF     & 11.80 & 13.26 & 15.55 & 16.82 & 27.17 & 29.32 & 31.78 & 32.31 \\
 & CensNet & 12.37 & 13.80 & 15.03 & 14.78 & 16.67 & 16.90 & 17.06 & 16.95 \\
 & {\bf HMGAT} & 16.74 & 18.03 & 20.60 & 21.05 & 30.23 & 32.19 & 33.99 & 33.97 \\
\hline
\end{tabular}
\label{table:HMGAT:users}
\end{table}

\subsubsection{Scalability across Different User Numbers}
Table~\ref{table:HMGAT:users} evaluates the scalability of HMGAT and CensNet under different numbers of users. We train models under $(8,16)$ and $(16,32)$ configurations and test them under $K\in\{6,7,8,9,10\}$ and $K\in\{14,15,16,17,18\}$. As observed, the HMGAT  demonstrates the best scalability across all scenarios. For example, in LA with $(16,32)$, the HMGAT achieves sum rates of 35.71, 37.03, 37.61, 38.35, and 39.03 for $K\in\{14,15,16,17,18\}$, compared to CensNet's sum rates of 27.88, 28.60, 29.03, 29.63, and 29.04. Meanwhile, the traditional methods (i.e., AO and PZF) performs significantly worse. Similarly, in CHI with $(16,32)$, the HMGAT maintains a stable growth in sum rate from 30.23 to 33.97 as the number of users increases, highlighting its superior performance  in large-scale user and antenna configurations.

\begin{figure}[h]
{\centering
{\includegraphics[ width=.45\textwidth]{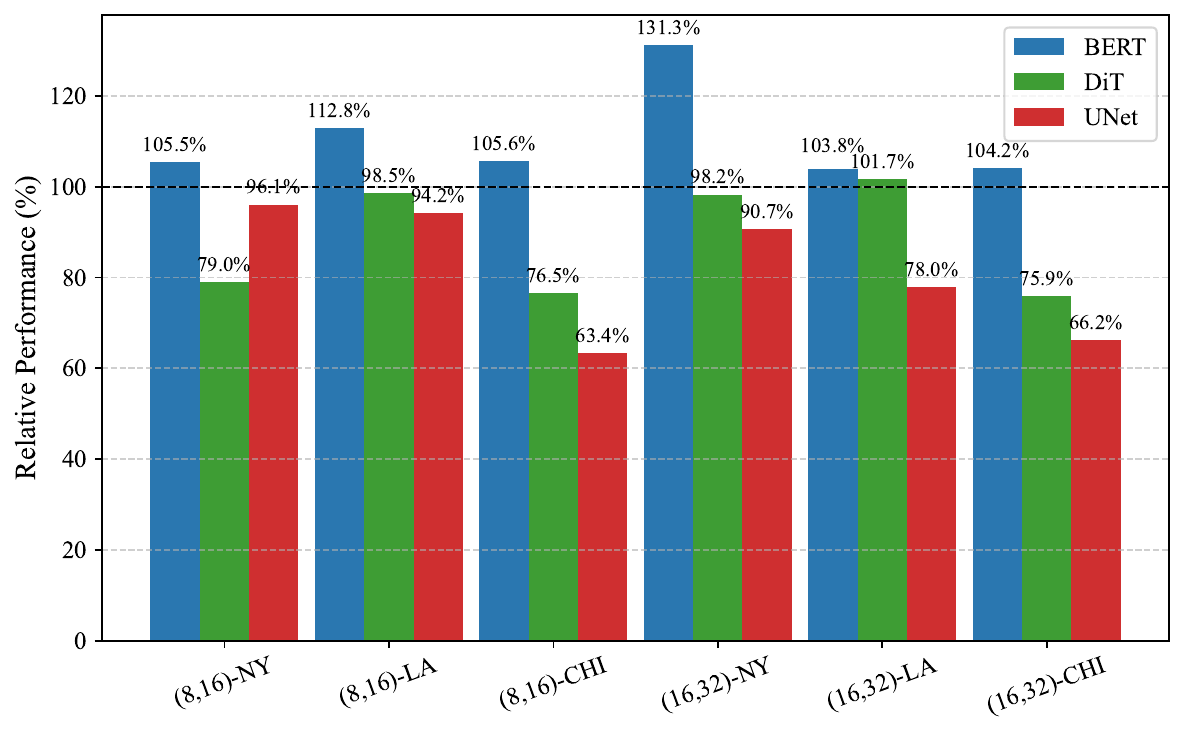}}}
\caption{Relative performance after augmentation.} 
\label{fig:BERT:aug}
\end{figure}

\subsection{Effectiveness of BERT-based CSI Generation}
To demonstrate the effectiveness of the proposed BERT-based NCSN, we generate 5,000 high-resolution CSI samples using the model. For comparison, we also employ UNet- and DiT-based NCSNs to generate CSI samples under the same experimental settings. These samples are employed in data augmentation experiments for enhancing the HMGAT. Besides, the similarity between the distributions of generated samples and original samples is analyzed.

\subsubsection{Data Augmentation} Fig.~\ref{fig:BERT:aug} evaluates the data augmentation performance of BERT-based NCSN in the comparison with  two baselines (i.e., UNet and DiT).  The generated CSI samples are added into training set of the HMGAT. It is observed that the BERT-based NCSN as the data augmentation module effectively enhances HMGAT, with improvements ranging from $3.8\%$ to $31.3\%$. In contrast, UNet and DiT fail to improve the performance of HMGAT. Particularly  in CHI with (16,32), the baselines incur  performance degradation of $24.1\%$ and $33.8\%$, respectively. The reason for the performance degradation could be that the generated CSI samples are not  i.i.d. as the original CSI samples, and results in overfitting of HMGAT.

\begin{table}[h]
\centering
\caption{KS statistics between original and generated CSI samples.}
\setlength{\tabcolsep}{6pt} 
\begin{tabular}{c|c|c|c|c}
\hline
$(K, N_{\rm T})$ & Model & NY & LA & CHI \\
\hline
\hline
\multirow{3}{*}{(8,16)}
  & BERT & 0.117 & 0.108 & 0.286     \\
  & DiT  & 0.350  & 0.735 & 0.543 \\
  & UNet & 0.732     & 0.770 & 0.938 \\
\hline
\multirow{3}{*}{(16,32)}
  & BERT & 0.268 & 0.153 & 0.080 \\
  & DiT  & 0.640 & 0.700 & 0.841 \\
  & UNet & 0.836 & 0.980  & 0.998 \\
\hline
\end{tabular}
\label{tab:BERT:KS}
\end{table}

\begin{table*}[htbp]
\centering
\caption{JS divergence of generated CSI components.}
\setlength{\tabcolsep}{3pt} 
\begin{tabular}{c|c||ccc|ccc|ccc}
\hline
\multirow{2}{*}{$(K, N_{\rm T})$} & \multirow{2}{*}{Model} 
& \multicolumn{3}{c|}{NY} & \multicolumn{3}{c|}{LA} & \multicolumn{3}{c}{CHI} \\
\cline{3-11}
 & & Real & Imaginary & Magnitude & Real & Imaginary & Magnitude & Real & Imaginary & Magnitude \\
\hline
\hline
\multirow{3}{*}{(8,16)} 
  & BERT  & $1.53 \mathrm{e}{-2}$ & $4.44 \mathrm{e}{-2}$ & $1.15 \mathrm{e}{-1}$ & $5.07 \mathrm{e}{-3}$ & $5.34 \mathrm{e}{-3}$ & $1.51 \mathrm{e}{-2}$ & $7.56 \mathrm{e}{-2}$ & $8.92 \mathrm{e}{-2}$ & $2.13 \mathrm{e}{-1}$ \\
  & DiT   & $3.19 \mathrm{e}{-2}$ & $6.71 \mathrm{e}{-2}$ & $1.76 \mathrm{e}{-1}$ & $8.68 \mathrm{e}{-2}$ & $8.53 \mathrm{e}{-2}$ &  $2.49 \mathrm{e}{-1}$ & $8.57 \mathrm{e}{-2}$ & $9.53 \mathrm{e}{-2}$ & $2.34 \mathrm{e}{-1}$ \\
  & UNet  & $5.78 \mathrm{e}{-2}$ & $8.67 \mathrm{e}{-2}$ & $1.86 \mathrm{e}{-1}$ & $1.53 \mathrm{e}{-1}$ & $1.48 \mathrm{e}{-1}$ & $2.79 \mathrm{e}{-1}$ & $1.74 \mathrm{e}{-1}$ & $1.90 \mathrm{e}{-1}$ & $3.54 \mathrm{e}{-1}$ \\
\hline
\multirow{3}{*}{(16,32)} 
  & BERT  & $1.81 \mathrm{e}{-2}$ & $3.54 \mathrm{e}{-2}$ & $1.05 \mathrm{e}{-1}$ & $5.06 \mathrm{e}{-3}$ & $3.82 \mathrm{e}{-3}$ & $1.37 \mathrm{e}{-2}$ & $7.96 \mathrm{e}{-2}$ & $9.07 \mathrm{e}{-2}$ & $1.99 \mathrm{e}{-1}$ \\
  & DiT   & $2.06 \mathrm{e}{-2}$ & $4.07 \mathrm{e}{-2}$ & $1.31 \mathrm{e}{-1}$ & $3.35 \mathrm{e}{-2}$ & $3.26 \mathrm{e}{-2}$ & $6.44 \mathrm{e}{-2}$ & $1.17 \mathrm{e}{-1}$ & $1.33 \mathrm{e}{-1}$ & $2.79 \mathrm{e}{-1}$ \\
  & UNet  & $4.99 \mathrm{e}{-2}$ & $6.82 \mathrm{e}{-2}$ & $1.76 \mathrm{e}{-1}$ & $3.13 \mathrm{e}{-1}$ & $3.12 \mathrm{e}{-1}$ & $3.47 \mathrm{e}{-1}$ & $1.80 \mathrm{e}{-1}$ & $1.93 \mathrm{e}{-1}$ & $3.57 \mathrm{e}{-1}$ \\
\hline
\end{tabular}
\label{table:BERT:JS}
\end{table*}

\subsubsection{Distribution Analysis of Generated CSI Samples}
We first evaluate the quality of the generated CSI from a performance perspective by comparing the performance distributions of the original and generated CSI under the same HBF algorithm. The two-sample Kolmogorov-Smirnov (KS) test is adopted to quantify the distributional similarity. As summarized in Table~\ref{tab:BERT:KS}, the CSI samples generated by BERT exhibit the closest match to the original data, with the minimum and maximum KS statistics being 0.08 and 0.286, respectively. In contrast, the samples generated by UNet and DiT show substantially larger deviations, particularly in CHI with (16,32), where all KS statistics exceed 0.8. These results indicate that BERT better preserves the performance-related characteristics of the original CSI.

To further examine the statistical fidelity of the generated CSI, we analyze the distributions of their real, imaginary, and magnitude components and compare them with the original CSI using the Jensen-Shannon (JS) divergence. The results in Table~\ref{table:BERT:JS} show that BERT consistently achieves the smallest distributional divergence across all scenarios, whereas the baselines produce significantly larger discrepancies. This further confirms that BERT effectively preserves the statistical properties of the original CSI and is therefore well suited for data augmentation.

\subsection{Effectiveness of DeBERT for  Denoising Imperfect CSI}
\begin{figure*}[htbp]
	\centering
	\begin{subfigure}{0.32\linewidth}
		\centering
    \includegraphics[width=1\linewidth]{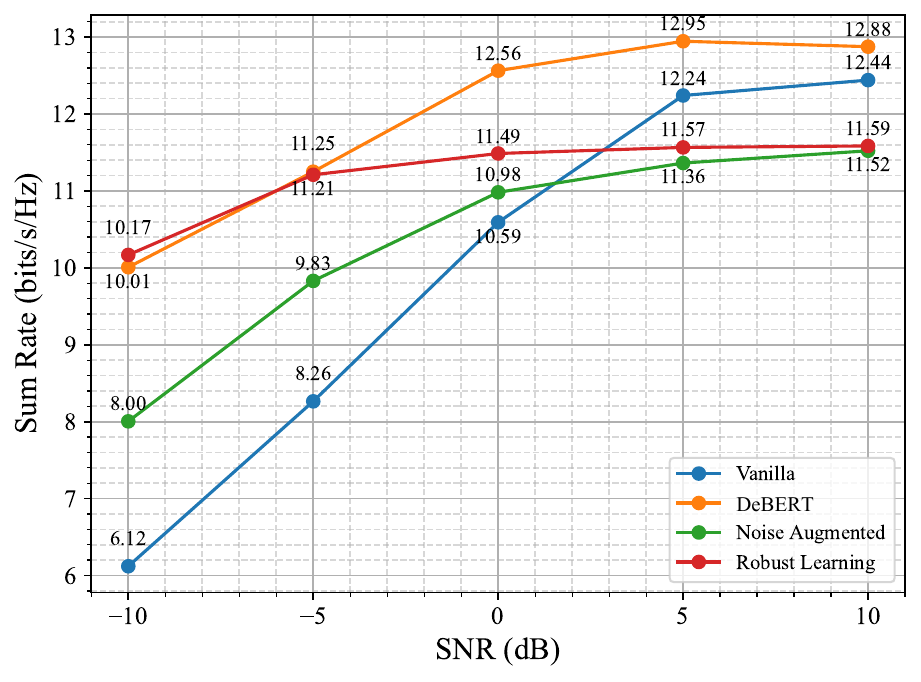}
		\caption{{NY $(K = 8,N_{\rm T} = 16)$}}
		\label{bs1}
	\end{subfigure}
	\centering
	\begin{subfigure}{0.32\linewidth}
		\centering
		\includegraphics[width=1\linewidth]{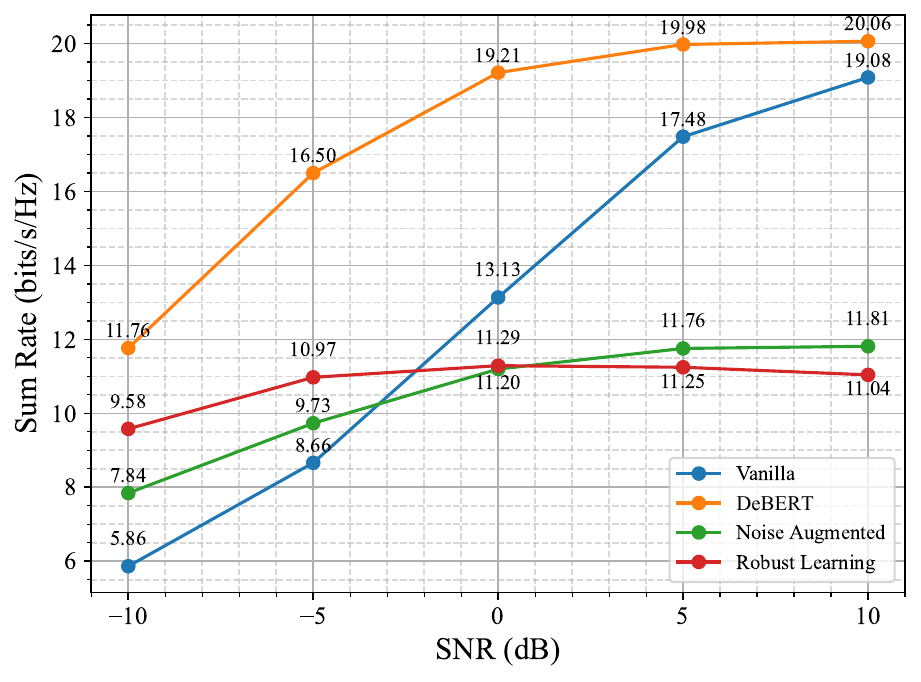}
		\caption{{LA $(K = 8,N_{\rm T} = 16)$}}
		\label{bs2}
	\end{subfigure}
	\centering
	\begin{subfigure}{0.32\linewidth}
		\centering
		\includegraphics[width=1\linewidth]{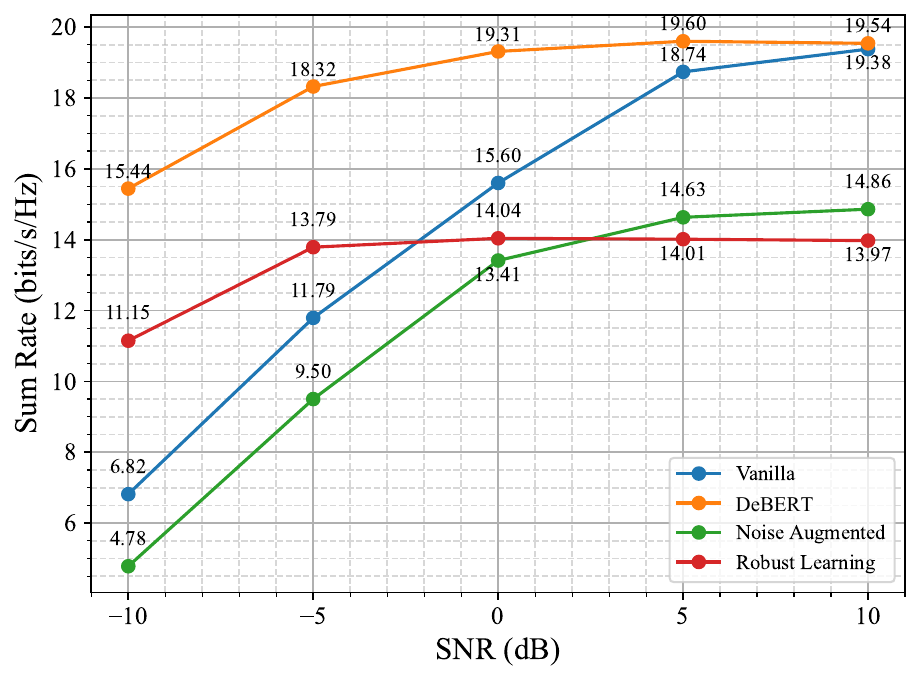}
		\caption{{CHI $(K = 8,N_{\rm T} = 16)$}}
		\label{bs3}
	\end{subfigure}

	\begin{subfigure}{0.32\linewidth}
		\centering
    	\includegraphics[width=\linewidth]{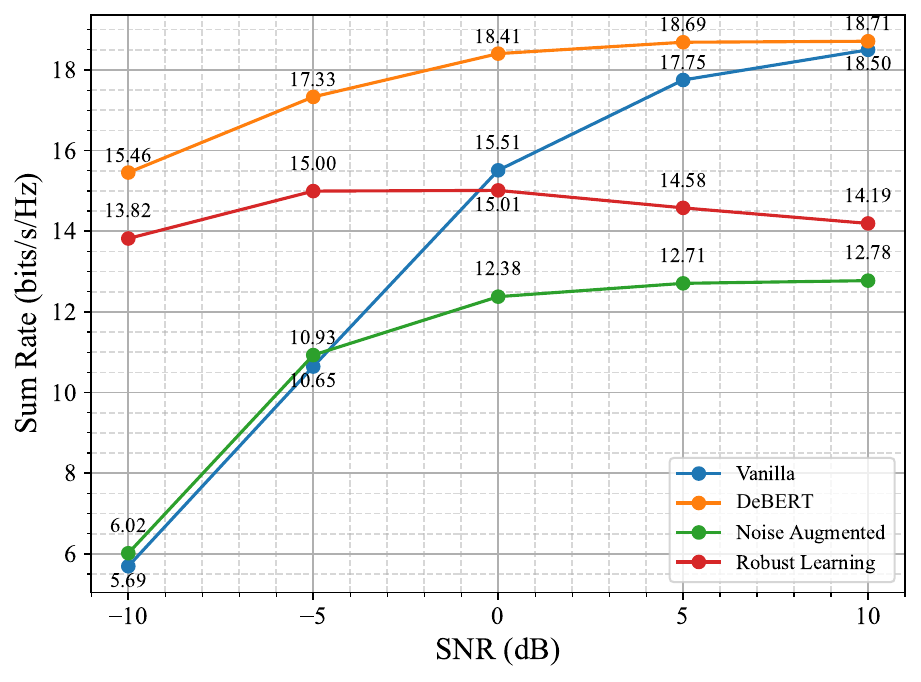}
		\caption{{NY  $(K = 16,N_{\rm T} = 32)$}}
		\label{bs4}
	\end{subfigure}
	\begin{subfigure}{0.32\linewidth}
		\centering
		\includegraphics[width=\linewidth]{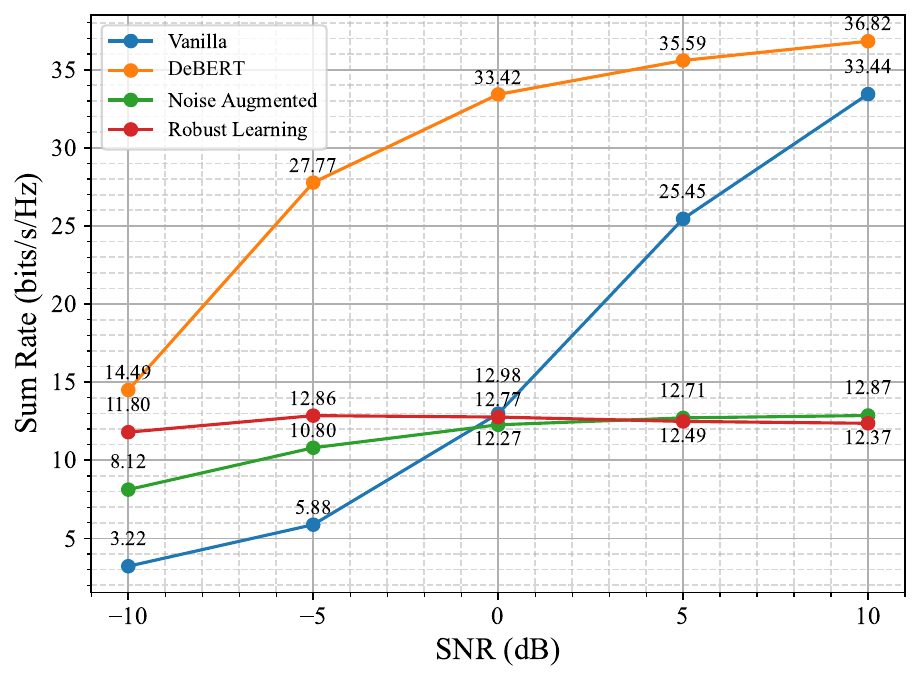}
		\caption{{LA $(K = 16,N_{\rm T} = 32)$}}
		\label{bs5}
	\end{subfigure}
	\begin{subfigure}{0.32\linewidth}
		\centering
		\includegraphics[width=\linewidth]{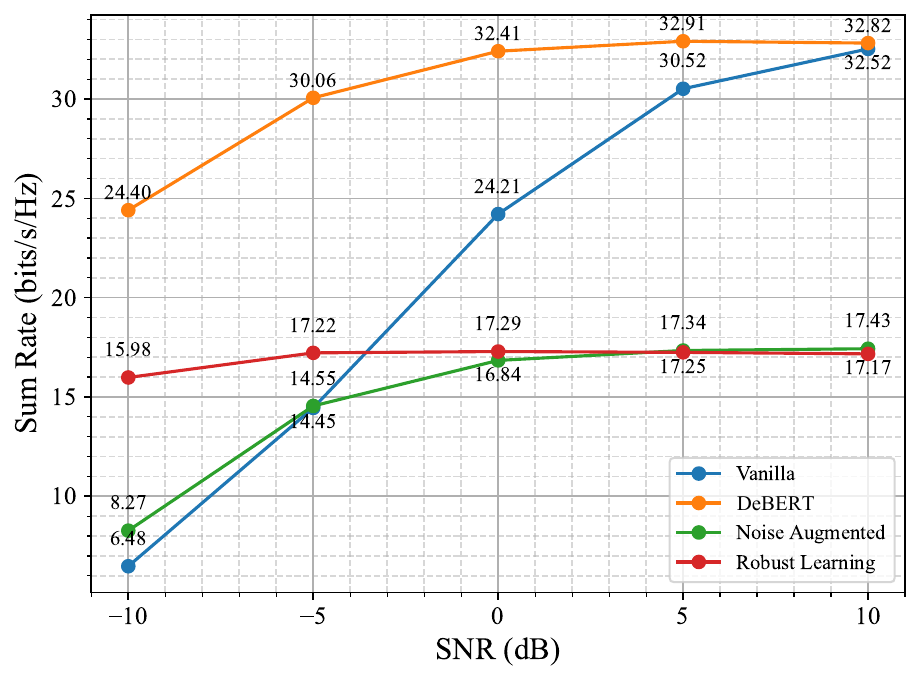}
		\caption{{CHI $(K = 16,N_{\rm T} = 32)$}}
		\label{bs6}
	\end{subfigure}
	\caption{Sum-rate performance under different channel error levels.}
	\label{fig:DeBERT:sumrate}
\end{figure*}

To evaluate the effectiveness of DeBERT for denoising imperfect CSI, we conduct experiments across multiple cities and antenna configurations under different CSI error levels $\delta_{\rm E}^2\in\{-10, -5, 0, 5, 10\}$ {dB}, as shown in Fig.~ \ref{fig:DeBERT:sumrate}. The results indicate that the vanilla model retains a certain  robustness under low-error conditions (e.g., 10 dB), but its performance deteriorates significantly as the  error level increases. In comparison, the Noise-Augmented and Robust Training methods, which enhance model robustness directly through training, effectively improve performance under high-error conditions (e.g., -10 dB and 5 dB). However, their performance degrades under low-error scenarios (e.g., 5 dB and 10 dB), revealing limitations in their applicability. Notably, the proposed DeBERT maintains strong performance across the entire range of error levels, demonstrating its effectiveness and broad applicability in various imperfect CSI environments.

\begin{figure}[htbp]
	\centering
	\begin{subfigure}{1\linewidth}
		\centering
    \includegraphics[width=1\linewidth]{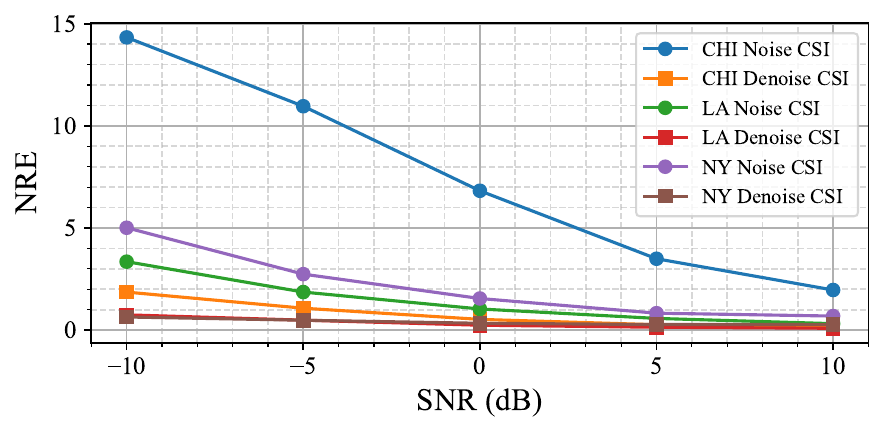}
		\caption{$(K = 8, N_{\rm T} = 16)$}
		\label{bs1}
	\end{subfigure}
    
	\begin{subfigure}{1\linewidth}
		\centering
    	\includegraphics[width=\linewidth]{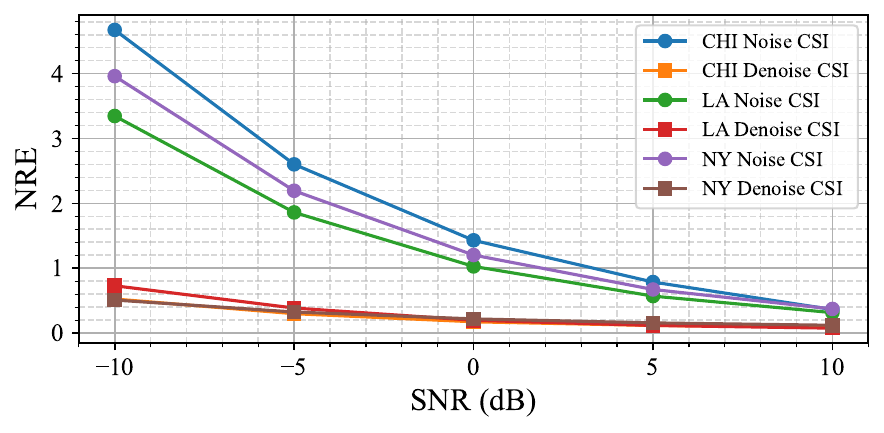}
		\caption{{$(K = 16, N_{\rm T} = 32)$}}
		\label{bs4}
	\end{subfigure}
	\caption{NRE performance of DeBERT for channel denoising.}
	\label{fig:DeBERT:NER}
\end{figure}

To further quantify the denoising capability of DeBERT, Fig.~\ref{fig:DeBERT:NER} evaluates the Normalized Reconstruction Error (NRE) under different error levels, defined as follows:
\begin{flalign}
\mathrm{NRE} = \frac{\|\mathbf{H} - \widehat{\mathbf{H}}\|_F}{\|\mathbf{H}\|_F}.
\end{flalign}
As observed, the DeBERT effectively denoises the imperfect CSI across all error conditions, recovering channel matrices that closely approximate the high-resolution CSI, thereby demonstrating its superior denoising performance.

\section{Conclusion}
In this paper, we have proposed to utilize GNNs and score-based generative models to address the HBF tasks under limited training data and imperfect CSI conditions. We have designed the HMGAT for sum-rate HBF, which explicitly incorporated node- and edge-level message passing, thereby enhancing the model's representation capacity. For data augmentation, we have proposed a BERT-based DCSN to generate high-resolution CSI samples; for CSI denoising, we have proposed a BERT-based DSN, i.e., DeBERT, to refine imperfect CSI under arbitrary channel error levels.  We have conducted extensive experiments on  DeepMIMO urban datasets to demonstrate that the proposed models can significantly enhance expressiveness and robustness. Our future works will explore extensions to multi-cell systems and extend the DCSN and DSN frameworks to other DL-based wireless communication  designs.

\normalem
\bibliographystyle{IEEEtran}
\bibliography{IEEEabrv,ref}

\end{document}